\documentclass[a4paper,11pt]{article}
\pdfoutput=1 
\usepackage{subcaption} 
\usepackage{jcappub, caption}
\usepackage[T1]{fontenc} 
\newcommand{\ud}{\mathrm{d}}    
\newcommand{\ue}{\mathrm{e}}
\newcommand{\ub}{\mathrm{b}}

\newcommand{\uh}{\mathrm{h}}
\newcommand{\ut}{\mathrm{t}}
\newcommand{\uk}{\mathrm{k}}
\newcommand{\us}{\mathrm{s}}

\usepackage{multirow, siunitx}
\usepackage{float}
\DeclareSIUnit\parsec{pc}
\usepackage{tikz,xcolor,hyperref,upgreek}

\definecolor{lime}{HTML}{A6CE39}
\DeclareRobustCommand{\orcidicon}{\hspace{-3mm}
	\begin{tikzpicture}
	\draw[lime, fill=lime] (0,0) 
	circle [radius=0.16] 
	node[white] {\hspace{0.1mm}{\fontfamily{qag}\selectfont \tiny ID}};
	\draw[white, fill=white] (-0.07,0.1) 
	circle [radius=0.01];
	\end{tikzpicture}
	\hspace{-5mm}
}

\foreach \x in {A, ..., Z}{\expandafter\xdef\csname orcid\x\endcsname{\noexpand\href{https://orcid.org/\csname orcidauthor\x\endcsname}
		{\noexpand\orcidicon}}
}

\title{Implications of \textit{SARAS3} data for Coulomb-like interacting dark matter}

\author[a,b,c,d,1]{Shikhar Mittal\orcidB{},}
\author[e,f,1]{Prakhar Bansal\note{Equal first authors}\orcidA{},}
\author[b,c]{Harry Bevins\orcidC{},}
\author[g]{and Saurabh Singh\orcidD{}}

\affiliation[a]{Manipal Institute of Technology Bengaluru, Manipal Academy of Higher Education, Manipal, India}
\affiliation[b]{Battcock Centre for Experimental Astrophysics, Cavendish Laboratory, J.~J.\ Thomson Avenue, Cambridge CB3 0HE, UK}
\affiliation[c]{Kavli Institute for Cosmology, University of Cambridge, Madingley Road, Cambridge CB3 0HA, UK}
\affiliation[d]{Tata Institute of Fundamental Research, Homi Bhabha Road, Mumbai 400005, India}
\affiliation[e]{Department of Physics and Leinweber Institute for Theoretical Physics, University of Michigan, 450 Church St, Ann Arbor, MI 48109, USA}
\affiliation[f]{Indian Institute of Technology Bombay, Mumbai 400076, India}
\affiliation[g]{Raman Research Institute, C. V. Raman Avenue, 5th Cross Road, Bangalore 560080, India}

\emailAdd{shikhar.mittal4@gmail.com}
\emailAdd{prakharb@umich.edu}
\emailAdd{h.bevins@imperial.ac.uk}
\emailAdd{saurabhs@rri.res.in}

\abstract{The 21-cm signal from cosmic dawn is a potentially sensitive probe of interactions between dark matter (DM) and baryons. We investigate the implications of the \textit{SARAS3} non-detection in the \SIrange{55.5}{84.4}{\mega\hertz} band for Coulomb-like interacting DM (IDM). In contrast to earlier constraint analyses that focused primarily on baryon cooling, we model the interaction self-consistently by including both excess cooling of the gas and the suppression of structure formation, which delays the onset of star formation and hence suppresses the Ly$\upalpha$, X-ray, and ionizing backgrounds at early times. We perform a joint Bayesian fit of a global 21-cm signal model and a flexible foreground model to the \textit{SARAS3} antenna temperature, and find that the signal parameters remain weakly constrained after marginalizing over the foregrounds. The null result is nonetheless informative: the data disfavour deep absorption features within the observed band, with the strongest bound at $z = 23.6$ ($\nu \approx \SI{57.7}{\mega\hertz}$), where $T_{21} \gtrsim -277.6\,$mK at $3\sigma$. Comparing the IDM and standard cold dark matter scenarios, we find no statistically significant preference for IDM (Bayes factor $B \approx 1.7$). While we do not constrain the strength of baryon–DM interactions, the \textit{SARAS3} non-detection places a meaningful upper bound on the amplitude of the global 21-cm signal in this class of models.}

\begin{document}
\maketitle
\flushbottom

\section{Introduction}\label{sec:intro}
We understand only $\lesssim5\,\%$ content of our Universe, which is in the form of baryonic matter that we see around us. The bulk of the matter content of the universe is invisible to us, and is known as dark matter (DM). DM is about five times more abundant than the baryonic matter, and evidence for its existence has been shown by several observations in the past decades \cite[e.g.,][]{Rubin_1970, Albada_1985, Trimble_1987, Clowe_2006}. The currently accepted model assumes that DM is cold (consequently abbreviated as `CDM') and collisionless. While CDM model has been quite successful in explaining a number of cosmological observations such as the large-scale structure \citep{Peebles_1980, Peacock_2001, Massey_2007}, there are some observations which hint towards non-gravitational DM and baryon interactions. For example, the universal relations between dynamical mass and baryonic mass of galaxies or galaxy clusters \citep{Chan_2022}, tension between $\Lambda$CDM-based cosmological simulations and observations of dwarf galaxies \citep{Salucci_2012, Papastergis_2015, Oman_2015}, anomalous positron abundance in the cosmic radiation \citep{Adriani_2009}, and surplus of gamma radiation coming from the centre of the Milky Way \citep{Christoph_2012}.

A major goal of particle physics is to discover any possible non-gravitational interactions of DM, and cosmological observations are but one avenue to probe such interactions~\cite[see e.g.,][]{Spergel_2000, Feng_2010, Manuel_2022, Slatyer_2024}. See \cite{Bergstrom_2009, Bertone_2018, Mahmoudi_2021} for a detailed list of evidence for DM, a discussion of DM candidates and possible detection strategies.

One proposed subclass of interacting DM particles is millicharged DM. As the name suggests, these hypothetical particles carry a tiny fractional electric charge and interact with charged particles. The strength of these interactions depends on the relative velocity between the DM particle and the baryonic particle \cite{Munoz_2015, Millicharge_2, Millicharge_3, Millicharge_4}. In the present work, we focus on a broader class of DM models featuring a Coulomb-like interaction for which the interaction cross-section scales inversely with the fourth power of the relative velocity due to elastic scattering \cite{Dvorkin_2014}. Unlike millicharged particles, Coulomb-like DM particles would interact with all baryons including neutral baryons. When required, in order to explicitly distinguish between the scenarios with a gravitational-interaction-only DM and the Coulomb-like interacting DM, we will use the term standard cold DM (or simply CDM for short), and IDM, respectively to refer to these specific scenarios.

Observations of large scale cosmic structure tell us that DM must have been much colder than the baryonic matter through most of the cosmic history of the Universe \cite{Turner_2000}. For standard CDM models, DM decouples with ordinary matter very early and henceforth the DM temperature is expected to follow adiabatic cooling, driven by the expansion of the Universe so that $T\sim1/a^2$. Baryons on the other hand, remain kinetically coupled to the cosmic microwave background (CMB) due to the presence of a small number of free electrons, till fairly late times $z\sim150$ \cite{Seager_1999, Seager_2000}. If, on the other hand, DM interacts with baryons via Coulomb-like interactions, it has interesting consequences for both the DM temperature, as well as the thermal history of intergalactic medium (IGM), i.e., the baryons. The interactions provide a channel to exchange heat with the baryonic matter, resulting in an excess cooling of baryonic matter, and a consequent heating of the DM, relative to the standard scenario.

Cosmology using the 21-cm signal~\cite{MMR, Furlanetto, Pritchard_2012} has the potential to reveal insights into the nature of DM from high redshifts in ways that no other current probes can \cite{Katz_2024}. Unfortunately, observations of high-redshift 21-cm signal are hindered by several systematics such as galactic foregrounds \cite{condon}, extragalactic point sources \cite{Mittal_eps}, instrumental systematics \cite{Scheutwinkel_2022}, ionospheric distortions \cite{Tripathi}, among others. Nevertheless, the first claimed detection by the \textit{EDGES} experiment has created a wider interest in this field \cite{EDGES_2018}. The stronger-than-expected cosmic dawn 21-cm absorption signal reported by the \textit{EDGES} collaboration sparked debate on its validity, e.g., challenging the foreground modelling \cite{Hills, Singh_2019}, unaccounted for ground plane systematics \cite{Bradley_2019} or calibration systematics \cite{Sims_2019}. Despite the conflicting views on the cosmological origin of the signal reported by the \textit{EDGES} collaboration, their result has inspired interesting literature on various cosmological and astrophysical processes that affect the physics of IGM (see, e.g., \cite{xiao_2019, Upala, Dhuria_2021, Mathur_2022, Mittal_jwst}, and references therein). In particular, models in which DM interacts with baryons received much attention \cite{Barkana_2018, Berlin_2018, Fialkov_2018, Slatyer_2018, Kovetz_Millicharged, Liu_2019, Barkana_2018_2} because such models can provide the necessary excess cooling of baryonic matter beyond the adiabatic cooling limit in order to explain strong \textit{EDGES} 21-cm signal.

In 2022, the Shaped Antenna Measurement of the Background Radio Spectrum (\textit{SARAS}) \cite{SARAS_Design} global 21-cm experiment published their independent findings. \textit{SARAS} uses an antenna deployed on water, unlike \textit{EDGES}, which is ground-based. This setup provides a homogeneous medium of high dielectric constant beneath the antenna, enhancing sensitivity and reducing radio wave interference from the ground. The 2022 data release \cite{SARAS_Detection} (hereafter referred to as \textit{SARAS3} data) rejected the presence of the \textit{EDGES} absorption profile at 95\,\% confidence. Upcoming data from global 21-cm experiments such as the Radio Experiment for the Analysis of Cosmic Hydrogen (\textit{REACH})\footnote{\url{https://www.reachtelescope.org/}} \cite{reach} could potentially resolve the disagreement between the two experiments.

Driskell et al.~\cite{Driskell} showed that an interacting DM model leads to two distinct effects on the 21-cm signal: cooling of ordinary baryons via energy exchange, \textit{and} delayed star formation due to suppression of halo formation through momentum exchange between DM and baryons. Previous works deriving constraints on millicharged dark matter \cite{Berlin_2018, Liu_2019, Barkana_2018_2} accounted only for the baryon-cooling effect and relied on the (now contested) \textit{EDGES} detection. Recent forecasts by Rahimieh et al.~\cite{Rahimieh_2025, Rahimieh_2026} examine the sensitivity of future global and interferometric 21-cm experiments as probes of baryon-DM interactions.

In this work, we derive the first constraints on a consistent IDM model -- one that incorporates both the baryon-cooling and the delayed-star-formation effects -- using the \textit{SARAS3} non-detection of the global 21-cm signal. We carry out a fully Bayesian statistical analysis, in contrast to the more approximate treatments adopted in previous constraint work. Given the noise level of the \textit{SARAS3} data, the resulting constraints on IDM and astrophysical parameters are modest; nonetheless, the analysis places upper limits on the amplitude of the global 21-cm signal for this class of models and demonstrates the framework that future, more sensitive measurements can apply.

This paper is organised as follows. In section~\ref{sec:method}, we describe our interacting DM model and discuss its impact on the 21-cm signal. In section~\ref{sec:infer} we give the description of the data and outline the details of our inference procedure. In section~\ref{sec:rnd} we present the results from our Bayesian inference, discuss implications and any potential limitations. Finally, we conclude in section~\ref{sec:conc}. We use the following cosmological parameters: $\Omega_{\mathrm{m}}= 0.315$, $\Omega_{\mathrm{b}}=0.049$, $\Omega_\Lambda = 0.685$, $h=0.674$, $Y_{\mathrm{p}}=0.245$, $T_0=\SI{2.725}{\kelvin}$, $\sigma_8 = 0.811$ and $n_{\mathrm{s}} = 0.965$ \cite{Fixsen_2009, Planck}, where $T_0$ and $Y_{\mathrm{p}}$ are the CMB temperature measured today and primordial helium fraction by mass, respectively.

\section{Theory and Methods}\label{sec:method}
In this section, we discuss the modelling of the thermal and ionization history of the intergalactic medium (IGM) and the global 21-cm signal in brief. We also discuss the impact of Coulomb-like interacting DM (IDM) on the IGM, and hence the 21-cm signal, in detail.

\subsection{The global 21-cm signal}\label{sec:gs}
We briefly overview the modelling of the IGM and the global 21-cm signal. (Readers familiar with the 21-cm signal calculations may jump to section~\ref{sec:dmm}, where we discuss the implications of an interacting DM model for the 21-cm signal.) We use the analytical-prescription-based Python package \texttt{ECHO21}\footnote{\url{https://github.com/shikharmittal04/echo21/}} \cite{Mittal_echo} to model the global 21-cm signal. We give below a brief overview of the approach taken in \texttt{ECHO21}. 

The 21-cm signal is a measure of the distortion in the background created by the hyperfine transitions in the neutral hydrogen present in the intervening IGM. The global 21-cm signal measured as a differential brightness temperature is given as \cite{MMR},
\begin{equation}  
\label{eq:deltaT21}
T_{21} = 27 \Bar{x}_{\textsc{Hi}} \left(\frac{1-Y_{\mathrm{p}}}{0.76}\right)\left(\frac{\Omega_{\ub} h^2}{0.023}\right)\sqrt{\frac{0.15}{\Omega_{\mathrm{m}} h^2}\frac{1+z}{10}}\left(1-\frac{T_{\gamma}}{T_{\mathrm{s}}}\right)\text{mK}\,,
\end{equation}
where $\Bar{x}_{\textsc{Hi}}$ is the globally-averaged neutral hydrogen fraction, $z$ is the redshift, $T_{\gamma}=T_{\gamma0}(1+z)$ is the CMB temperature and $T_{\mathrm{s}}$ is the spin temperature. We account for the collisional coupling ($x_{\mathrm{k}}$) \cite{Mittal_pbh} and the Ly$\upalpha$ coupling ($x_{\mathrm{Ly}}$) \cite{Mittal_lya}, so that the spin temperature is given as
\begin{equation}
T_{\mathrm{s}}^{-1}= \frac{T_{\gamma}^{-1}+ (x_{\mathrm{k}}+x_{\mathrm{Ly}})T_{\mathrm{k}}^{-1}}{1+ x_{\mathrm{k}}+x_{\mathrm{Ly}} }, \, 
\end{equation}
where $T_{\mathrm{k}}$ is the gas kinetic temperature. Note that our background is just the CMB and does not include any excess radio background which might enhance the 21-cm signal \cite{Fialkov_19, Mittal_erb}.

We closely follow Mittal et al.~\cite{Mittal_lya} for modelling $x_{\mathrm{Ly}}$, except for the modelling of the star formation rate density (SFRD), which we discuss in the next section. We use Population-II stars for modelling the spectral energy distribution (SED) of Lyman series photons giving us our default model for $x_{\mathrm{Ly}}$ \cite{BL05}. To capture the high-redshift uncertainty in the SED modelling we use a dimensionless fudge factor $f_{\mathrm{Ly}}$ so that
\begin{equation}
J_{\mathrm{Ly}} \to f_{\mathrm{Ly}} J_{\mathrm{Ly}}\,,
\end{equation}
where $J_{\mathrm{Ly}}$ is the background specific intensity of cosmic Ly$\upalpha$ photons. As both, Ly$\upalpha$ coupling and heating ($q_{\mathrm{Ly}}$) are directly dependent on background specific intensity \cite{Mittal_lya}, $f_{\mathrm{Ly}}$ accordingly impacts $x_{\mathrm{Ly}}$ and $q_{\mathrm{Ly}}$, i.e.,
\begin{equation}
x_{\mathrm{Ly}},q_{\mathrm{Ly}}\propto f_{\mathrm{Ly}}J_{\mathrm{Ly}}\,.\label{eq:falpha_to_heating}
\end{equation}
The parameter $f_{\mathrm{Ly}}$ becomes one of the free parameters of our model. A typical value of $f_{\mathrm{Ly}}=1$ corresponds to 10,000 Lyman series photons per baryon. This parameter is unknown, but we will choose a range of $10^{-2}$ to $10^2$, which covers a wide range of SED amplitudes, sampled uniformly on a logarithmic scale, for our inference analysis described in section~\ref{sec:infer}. Note that we do not account for the multiple scattering of Ly$\upalpha$ photons \cite{Reis2021, Semelin_2023, Mittal_mcrt, Flitter_2026} for the computation of $J_{\mathrm{Ly}}$ in this work. Multiple scattering refers to the repeated resonant scattering of Ly$\alpha$ photons by neutral hydrogen atoms before the photons redshift out of the line. This process is neglected in simplified treatments, where photons are assumed to propagate freely until they scatter once at resonance. Previous work has shown that multiple scattering reduces the strength of the global 21-cm signal \cite{Mittal_mcrt}. The resulting suppression is approximately degenerate with the effect of varying the Ly$\alpha$ efficiency parameter $f_{\mathrm{Ly}}$. Consequently, the impact of multiple scattering can be largely absorbed into an effective rescaling of $f_{\mathrm{Ly}}$, and we do not expect its omission to qualitatively affect the astrophysical or IDM constraints derived in this work.

For the two-zone model of IGM, the bulk IGM and \textsc{H\,ii} regions can be assumed to evolve independently to a good approximation. The evolution of the bulk IGM free electron fraction ($x_{\text{e}}$) can be captured by the effective-three-level-atom recombination equation \cite{Seager_1999, Seager_2000},
\begin{equation}
(1+z)H\frac{\ud x_{\text{e}}}{\ud z}=\mathcal{C}_{\mathrm{P}}\left[x_{\text{e}}^2n_{\text{H}}\alpha_{\mathrm{B}}-\beta(1-x_{\text{e}})\ue^{-E_\upalpha/(k_\mathrm{B}T_{\gamma})}\right]-\Gamma_{\mathrm{X}}(1-x_{\text{e}})\,,\label{xe}
\end{equation}
where $H$ is the Hubble rate at redshift $z$, $E_\upalpha=\SI{10.2}{\electronvolt}$, $\mathcal{C}_{\mathrm{P}}$ is the Peebles' factor, $n_{\text{H}}$ is the proper hydrogen number density, $\alpha_{\mathrm{B}}$ is the case-B recombination coefficient evaluated at gas temperature, $\beta$ is the photoionization rate evaluated at CMB temperature \cite{Haimoud_2010, Chluba_2011, Chluba_2015}, and $\Gamma_{\mathrm{X}}$ is the ionization rate due to X-ray photons.

Coupled to the ionization equation is the equation governing the evolution of IGM gas temperature ($T_{\mathrm{k}}$) which can be written as 
\begin{equation}
(1+z)\frac{\ud T_{\mathrm{k}}}{\ud z}=2T_{\mathrm{k}}-\frac{T_{\mathrm{k}}(1+z)}{1+x_\mathrm{He}+x_\mathrm{e}}\frac{\ud x_\mathrm{e}}{\ud z}-\frac{2}{3n_{\mathrm{b}}k_{\mathrm{B}}H}\sum q\,,\label{DEofTk}
\end{equation}
where $n_{\ub}$ is the baryon (including electrons) number density, $x_{\mathrm{He}}=Y_{\mathrm{p}}/[4(1-Y_{\mathrm{p}})]$ is the helium number fraction and $q$ is the volumetric heating rate.

Throughout the dark ages and cosmic dawn we consider adiabatic cooling due to cosmological expansion, change in internal energy due to changing electron number, and Compton heating due to electron-photon interaction \cite{Weymann}. Once star formation commences, which we set to be at $z=59$ \cite{Mirocha_2019}, we also include the `X-ray heating' due to X-rays photons and `Ly$\upalpha$ heating' due to Ly$\upalpha$ photons. We assume that both, X-ray and Ly$\upalpha$ emissivity track the instantaneous SFRD. For X-rays, we adopt a simple parametric approach; we calibrate X-ray emissivity to the X-ray luminosity per unit star formation rate (SFR) of the local galaxy observations \cite{F06}. Adopting the value based on \textit{Chandra}'s observation of 88 nearby galaxies we have \cite{Lehmer_2024}
\begin{equation}
\frac{L_{\mathrm{X}}}{\mathrm{SFR}}=2.45\times10^{32}\,\frac{\si{\watt}}{\mathrm{M_{\odot}yr^{-1}}}\,,\label{eq:xray}
\end{equation}
where $L_{\mathrm{X}}$ represents the integrated X-ray luminosity corresponding to energy range of $0.5$ to $\SI{8}{\kilo\electronvolt}$. To account for the contribution of low- and high-energy photons we extrapolate $L_{\mathrm{X}}$ to energies ranging from $0.2$ to $\SI{30}{\kilo\electronvolt}$. For this we assume that the specific (monochromatic) luminosity $l_{\mathrm{X}}$ goes as $l_{\mathrm{X}}(E)\propto E^{-w}$, for a free parameter $w$, typically taken to be 1.5 \cite{Mesinger_11}.

We introduce a fudge factor $f_{\mathrm{X}}$, which encapsulates any high-redshift uncertainties embedded in $L_{\mathrm{X}}/\mathrm{SFR}$. Additionally, $f_{\mathrm{X}}$ also serves as a free parameter to scale the strength of the X-ray background. We calibrate $f_{\mathrm{X}}=1$ to eq.~\eqref{eq:xray}. We note that not all of the X-ray luminosity is effective in IGM heating. Following ref.~\cite{Shull}, we use an electron fraction-dependent empirical fit for the fraction of X-ray luminosity responsible for the heating of IGM. In addition to these standard contributions we add a non-standard heating rate in this work resulting from the baryon-Coulomb-like DM (hereafter b$\chi$) interaction. We discuss the physics of b$\chi$ interaction in the next section.

For the \textsc{H\,ii} region, we assume that they remain at a fixed temperature of $\SI{e4}{\kelvin}$. We track the volume fraction of these pockets of fully ionized regions, $Q$, by the analytical reionization equation \cite{Madau_1999}
\begin{equation}
(1+z)H\frac{\ud Q}{\ud z}=-\frac{\dot{n}_{\mathrm{ion}}}{n^0_{\mathrm{H}}}+(1+x_{\mathrm{He}})n_{\mathrm{H}}\alpha_{\mathrm{B},10^4}CQ\,,\label{eq:reion}
\end{equation}
where $\dot{n}_{\mathrm{ion}}$ is the galaxy-dominated comoving ionizing emissivity, $n^0_{\mathrm{H}}$ is the comoving number density of hydrogen, and $C$ is the clumping factor of ionized hydrogen \cite{Shull_2012}. To determine the emissivity, we assume that it tracks the instantaneous SFRD, $\dot{\rho}_\star$. This gives us
\begin{equation}
\dot{n}_{\mathrm{ion}}=f_{\mathrm{esc}}I_{\mathrm{ion}}\dot{\rho}_\star\,,
\end{equation}
where $I_{\mathrm{ion}}$ is the ionizing photon yield and $f_{\mathrm{esc}}$ is the fraction of ionizing photons that escape the galaxies and leak into the IGM. For the Kroupa IMF and a metallicity of $0.01Z_{\odot}$ we have \citep{Madau_2017}
\begin{equation}
I_{\mathrm{ion}}=10^{53.44}\,\mathrm{s}^{-1}\left(\mathrm{M}_{\odot}\,\mathrm{yr}^{-1}\right)^{-1}\,.
\end{equation}
Modelling of the SFRD is discussed in section~\ref{sec:delayedSFR}.

\subsection{Dark matter model}\label{sec:dmm}
As mentioned in the introduction `Coulomb-like' (or any other interacting) DM can exchange both heat and momentum with the baryons. Heat exchange leads to excess baryon cooling, while momentum exchange suppresses the growth of small-scale structure and delays star formation. These two-fold effects can alter the predictions for the 21-cm signal. We will describe the effects of energy exchange and momentum exchange separately in this section. We note that we consider all baryonic species, i.e, hydrogen (both, neutral and ionized), helium, as well as electrons, as the scattering target in this work following the convention adopted in several previous studies of baryon-DM interactions \cite{Munoz_2015, Barkana_2018, Fialkov_2018, Xu_2018, Driskell, Flitter_2024, Cang_2025}.

\subsubsection{Excess baryon cooling}
We first describe the relevant equations for the IDM that govern the heat exchange between DM and baryons \cite{Munoz_2015}. The cross-section for the b$\chi$ interactions is defined as 
\begin{equation}
   \sigma(v_{\ub\chi}) = {\sigma_0} \left(\frac{v_{\ub\chi}}{{c}}\right)^{-4}\,,
\end{equation}
where $v_{\ub\chi}$ is the relative velocity between DM and baryons, ${c}$ is the speed of light and ${\sigma_0}$ is a free parameter which controls the interaction cross-section between DM and baryons. For brevity of notation, we use $\sigma_{45}$ defined as

\begin{equation}
\sigma_{45} = \frac{\sigma_0}{\SI{e-45}{\metre^2}}\,.
\end{equation}

Baryon-DM interaction leads to heat exchange between the two species which leads to cooling of baryons and simultaneously heating of DM. The heating rate of baryons and DM are respectively given by (in dimensions of temperature per unit time) \cite{Munoz_2015}
\begin{equation}
\dot{Q}_\uk = \frac{2m_\ub\rho_\chi c^4\sigma_0\ue^{-{r^2_{\ub\chi}}/{2}}}{(m_\chi+m_\ub)^2\sqrt{2\pi}u^3_{\ut\uh}}(T_\chi-T_\uk) + \frac{1}{k_\mathrm{B}}\frac{\rho_\chi}{\rho_\chi+\rho_\ub}\frac{m_\chi m_\ub}{m_\chi+m_\ub}v_{\ub\chi}D(v_{\ub\chi})\,,\label{eq:Qk}
\end{equation}
and
\begin{equation}
\dot{Q}_\chi = \frac{2m_\chi\rho_\ub c^4\sigma_0\ue^{-{r^2_{\ub\chi}}/{2}}}{(m_\chi+m_\ub)^2\sqrt{2\pi}u^3_{\ut\uh}}(T_\uk-T_\chi) + \frac{1}{k_\mathrm{B}}\frac{\rho_\ub}{\rho_\ub+\rho_\chi}\frac{m_\chi m_\ub}{m_\chi+m_\ub}v_{\ub\chi}D(v_{\ub\chi})\,. \label{eq:Qd}
\end{equation}
 Here $\rho_\chi = f_{\mathrm{DM}}\rho_\ud$; $f_{\mathrm{DM}}$ is the fraction of total DM which is Coulomb-like and $\rho_\ud$ is the total DM mass density ($\si{\kilo\gram\metre^{-3}}$). If $\rho_{\mathrm{crit}}$ is the critical Universe density then $\rho_\ud=(\Omega_\mathrm{m}-\Omega_\mathrm{b})\rho_{\mathrm{crit}}$. In this work we assume that all DM particles are of the same kind so that $f_{\mathrm{DM}}=1$. $\rho_\ub=\Omega_\mathrm{b}\rho_{\mathrm{crit}}$ is the total baryonic mass density and $m_\chi$ is the DM particle mass. When we speak of baryons we also include electrons as the heat transferred to gas is also shared by electrons. Accordingly, a number-averaged baryon mass can be defined as
\begin{equation}
m_\ub(x_{\ue}) = \frac{m_\mathrm{H}}{1+ (m_\mathrm{H}/m_\mathrm{He}-1)Y_\mathrm{P}+(1-Y_\mathrm{P})x_\mathrm{e}}\,.
\end{equation}
As an example, when the Universe is fully neutral ($x_{\ue}=0$) such as before cosmic dawn, then we have $m_\ub= 1.22m_\mathrm{H}$.

The thermal sound speed of the b$\chi$ fluid is defined as
\begin{equation}
u_{\ut\uh} = \sqrt{\frac{k_\mathrm{B}T_\mathrm{k}}{m_\ub}+\frac{k_\mathrm{B}T_\chi}{m_\chi} }\,,
\end{equation}
where $T_\chi$ is the temperature of DM particles.

The relative velocity in addition to decaying with the Hubble expansion as $1/a$, is affected by `drag' arising due to b$\chi$ interaction. This drag results in an extra contribution to internal heating of both baryons and DM via friction (conversion of bulk flow energy to internal thermal energy). This drag term is given by (in units of $\si{\metre\second^{-2}}$)
\begin{equation}
D(v_{\ub\chi}) = \sigma_0c^4\frac{\rho_\chi + \rho_\ub}{m_\chi+m_\ub}\frac{F(r_{\ub\chi})}{v^2_{\ub\chi}}\,,\label{eq:drag}
\end{equation}
where the function $F(x)$ is defined as
\begin{equation}
F(x) = \text{erf}\left(\frac{x}{\sqrt{2}}\right)-\sqrt{\frac{2}{\pi}}x \ue^{-x^2/2}\,,
\end{equation}
where `erf' represents the standard error function. The ratio of b$\chi$ relative velocity to the thermal velocity appearing in eqs.~\eqref{eq:Qk}, \eqref{eq:Qd}, and \eqref{eq:drag} is
\begin{equation}
r_{\ub\chi} = \frac{v_{\ub\chi}}{u_{\ut\uh}}\,.
\end{equation}

The evolution equations of the heating rates and drag term are to be used in conjunction with the equations governing the evolution of temperature of IDM particles $T_\chi$ and the relative velocity between DM and baryons, $v_{\ub\chi}$. These equations are given by
\begin{equation}
(1+z)\frac{\ud T_{{\chi}}}{\ud z} = 2T_\chi - \frac{2\dot{Q}_\chi}{3H}\,,\label{eq:tx}
\end{equation}
and
\begin{equation}
(1+z)\frac{\ud v_{\ub{\chi}}}{\ud z} = v_{\ub\chi} +\frac{D(v_{\ub\chi})}{H}\,,\label{eq:vbx}
\end{equation}
respectively. The impact of heat exchange with the DM bath on the gas temperature $T_\uk$ must be taken into account by including a heating term $q_\uk = n_\ub k_\mathrm{B}\dot{Q}_\uk$ on the RHS of eq.~$\eqref{DEofTk}$.

We now have a system of four equations, \eqref{xe}, \eqref{DEofTk}, \eqref{eq:tx}, and \eqref{eq:vbx} which need to be solved simultaneously. We initialise the system at $z = 1500$, where we expect the Saha's equation (equilibrium version of eq.~\ref{xe}) to be valid. Accordingly, the initial conditions we obtain are $x_{\ue} = 0.949$ and $T_\mathrm{k} = T_\gamma (z = 1500)$. Baryon and DM temperatures, and hence the 21-cm signal, are insensitive to small changes in the initial DM temperature. Thus, one may choose $T_\chi = 0$ at $z=1500$ which is similar to the expectation of perfectly cold DM. This approximation should be valid before the effect of interactions with baryons kick in and raise the temperature of the DM. Also, in ref.~\cite{Munoz_2015} the system was initialised at $z=1010$ where $v_{\ub{\chi}} = {v_{\mathrm{RMS}}}=\SI{29}{\kilo\metre\,\second^{-1}}$. Assuming a $v_{\ub{\chi}}\propto (1+z)$ behaviour at high redshifts, we set $v_{\ub{\chi}} = \SI{43.5}{\kilo\metre\,\second^{-1}}$ at $z=1500$.

\begin{figure}
\centering
\includegraphics[width=0.80\textwidth]{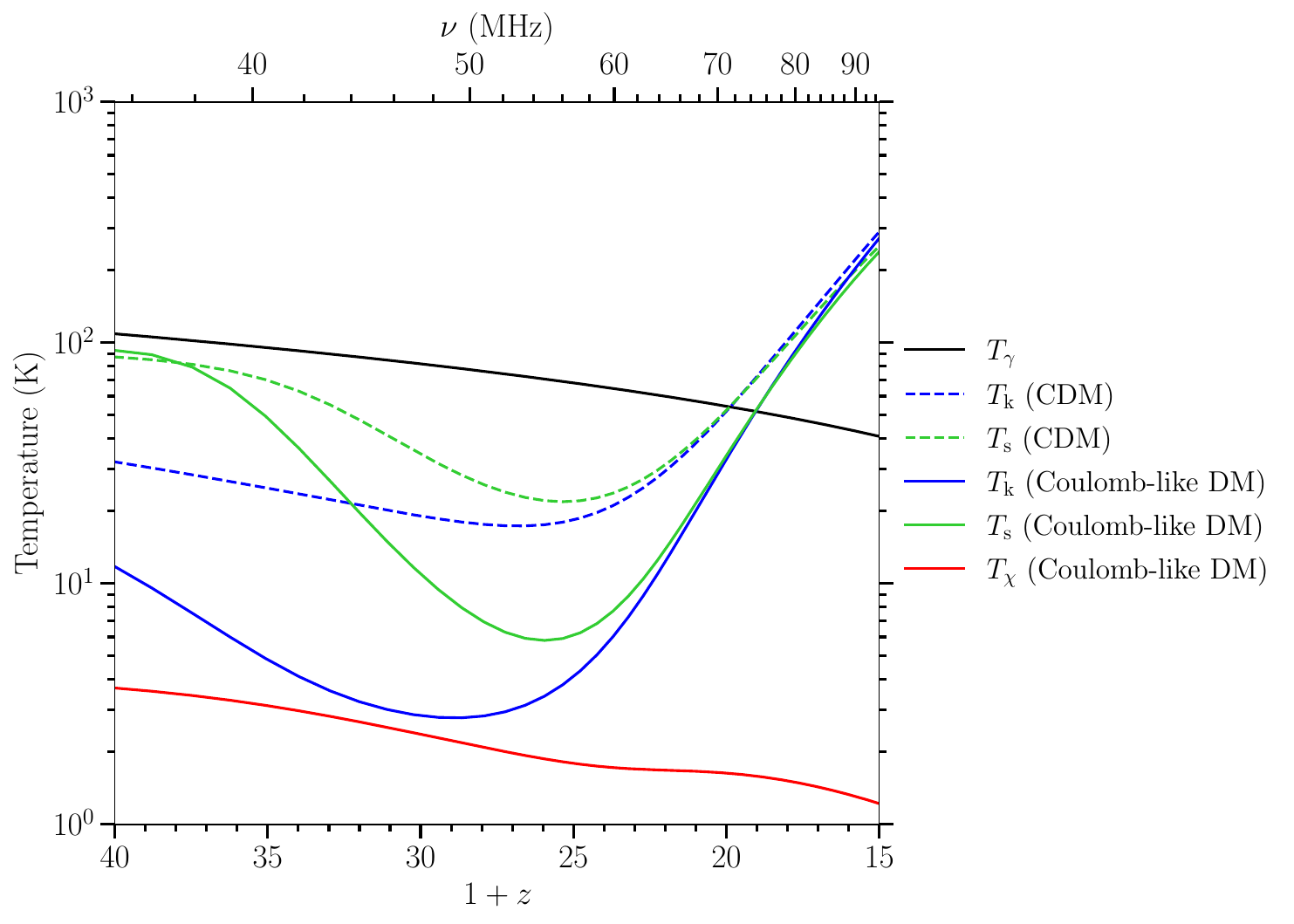}
\caption{Evolution of the CMB temperature ($T_\gamma$, black), gas temperature ($T_\uk$, blue), spin temperature ($T_\us$, green) and the IDM temperature ($T_\chi$, red). Solid curves correspond to DM while the dashed curves show the standard CDM case. The DM particle considered for the b$\chi$ system has a mass $m_\chi = \SI{0.32}{\giga\electronvolt}$ and cross-section $\sigma_{45} = 1$. The astrophysical parameters used are $f_{\mathrm{Ly}}=0.1, f_{\mathrm{X}} =1, w=1.5, f_{\mathrm{esc}}=0.01$,  min($T_{\text{vir}}) = \SI{500}{\kelvin}$. Note that we have considered only an excess cooling of baryons and no SFR suppression in this figure.}\label{fig:cooling}
\end{figure}

Figure~\ref{fig:cooling} shows a representative illustration of the evolution of the spin ($T_{\mathrm{s}}$), baryon ($T_{\uk}$), and DM ($T_{\chi}$) temperature for a system of IDM (solid curves). We choose $m_\chi = \SI{0.32}{\giga\electronvolt}$ and $\sigma_{45} = 1$ for IDM. For comparison we also show the corresponding curves for the standard CDM (dashed curves). As evident we see excess cooling of baryons (blue solid) compared to the case of standard CDM (blue dashed) as a result of interaction with DM. For redshifts $z>15$ we note that the temperature of DM-cooled baryons is more than an order of magnitude lower than the gas temperature in the standard adiabatically cooled IGM. In the b$\chi$ system, because we have a lower gas temperature, we have a lower spin temperature -- which in turn leads to a deeper absorption trough in the global 21-cm signal. The rise in gas temperature (and hence the spin temperature) in either case is predominantly due to X-ray heating \cite{Barkana_2018, Fialkov_2018}. Note that the delay in star formation due to b$\chi$ interaction reported in \cite{Driskell} is not yet accounted for in the results shown in figure~\ref{fig:cooling}. We discuss this effect in the next subsection.

\subsubsection{Delayed star formation}
\label{sec:delayedSFR}
As described in section \ref{sec:gs}, Ly$\upalpha$, X-ray, and ionizing emissivity track the star formation rate density (SFRD); these emissivities in turn decide the Ly$\upalpha$ coupling, Ly$\upalpha$ heating, X-ray heating, and ionizing rate. In this section we will describe how IDM affects the SFRD and hence the 21-cm signal. 

Driskell et al.~\cite{Driskell} pointed out that in addition to heat exchange there is also a momentum transfer between DM and baryons. The momentum transfer between DM and baryons alters the Boltzmann equations for the evolution of the DM and baryon overdensities and suppresses the matter power spectrum, $P(k)$, relative to the case of standard CDM, which in turn suppresses the mass variance $\sigma^2(M)$. The mass variance is the variance of the linear density field and is given by
\begin{equation}
\sigma^2(M)=\frac{1}{2\pi^2}\int_{0}^{\infty} k^2P(k)W(k|M)\,\ud k\,,
\end{equation}
where $W(k|M)$ is a window function, conventionally a top-hat filter. In this work, we use a different choice of a sharp-$k$ filter identical to that described in \cite{Driskell}. This choice of the window function is motivated by the inefficiency of the top-hat filter in modelling the power spectrum in modified cosmologies. When required at a general redshift, we extrapolate the above $\sigma$ to redshift $z$ using linear theory.

The mass variance gives the halo mass function (HMF) as follows
\begin{equation}
\frac{\ud n}{\ud M} = f(\sigma)\frac{\bar{\rho}^0_{\mathrm{m}}}{M}\frac{\ud \ln (\sigma^{-1})}{\ud M}\,,
\end{equation}
where $n$ is the comoving number density of haloes, $M$ is the halo mass, $\bar{\rho}^0_{\mathrm{m}}$ is the mean matter (DM and baryons) density today, and $f(\sigma)$ is a fitting function tuned to match the HMF predicted by $N$-body simulations. Throughout this work we use Tinker et al.~\cite{Tinker_2008} form for $f(\sigma)$.

Once we have obtained the HMF, we can compute the collapse fraction, $F_{\text{coll}}$ -- fraction of total matter that has collapsed into DM haloes more massive than $M_{\mathrm{min}}$ as
\begin{equation}
F_{\mathrm{coll}}=\frac{1}{\bar{\rho}^0_{\mathrm{m}}}\int_{M_{\mathrm{min}}}^{\infty} M\frac{\ud n}{\ud M}\,\ud M\,,
\end{equation}
where $M_\text{min}$ is the minimum halo mass capable of hosting a star. The redshift dependence of $F_{\mathrm{coll}}$ enters through $M_{\mathrm{min}}$ (see below) and $\sigma$.

The minimum halo mass $M_{\mathrm{min}}$ capable of hosting stars can be parametrized by the minimum virial temperature $T_{\mathrm{vir}}^{\mathrm{min}}$ as \citep{BL04, DAYAL20181},
\begin{equation}
M_{\mathrm{min}}=10^8\frac{1}{\sqrt{\Omega_{\mathrm{m}}h^2}}\mathrm{M}_{\odot}\left[\frac{10}{1+z}\frac{0.6}{m_\ub/m_\mathrm{H}}\frac{T^{\mathrm{min}}_{\mathrm{vir}}}{\num{1.98e4}}\right]^{3/2}\,,\label{tvir}
\end{equation}
where $h$ is the Hubble's constant measured today in units of $\SI{100}{\kilo\metre\per\second\per\mega\parsec}$. For the cosmic dawn period it is sufficient to take $m_\ub\approx1.22m_\mathrm{H}$, corresponding to the average baryon mass in a neutral universe.

We can now define our SFRD. The comoving SFRD, represented by $\dot{\rho}_\star(z)$, measured in units of mass per unit time per unit comoving volume, is set by the rate at which baryons collapse into DM haloes \citep{BL05}. Assuming that the collapse of baryons track the DM  overdensities, the SFRD will be proportional to the rate of collapse of DM haloes ($\ud F_{\mathrm{coll}}/\ud t$) and the baryon density (or density fraction), and hence the complete expression (in terms of $z$ derivative) is 
\begin{equation}
\dot{\rho}_\star(z)=-f_\star\cdot\bar{\rho}_\mathrm{b}^0\cdot (1+z)\cdot H(z)\cdot\frac{\ud F_{\mathrm{coll}}(z)}{\ud z}\,,\label{fstar}
\end{equation}
where $\bar{\rho}_\mathrm{b}^0$ is the mean cosmic baryonic mass density measured today, $f_\star$ is the star formation efficiency (SFE), defined as the fraction of baryons converted into stars in the star forming haloes. As far as the prediction of 21-cm signal is concerned, $f_\star$ is degenerate with $f_{\mathrm{Ly}}$, $f_{\mathrm{X}}$, and $f_{\mathrm{esc}}$. Accordingly, we do not consider $f_\star$ as a free parameter; we set $f_\star=0.1$ throughout this work.

\begin{figure}
\centering
\includegraphics[width=0.5\textwidth]{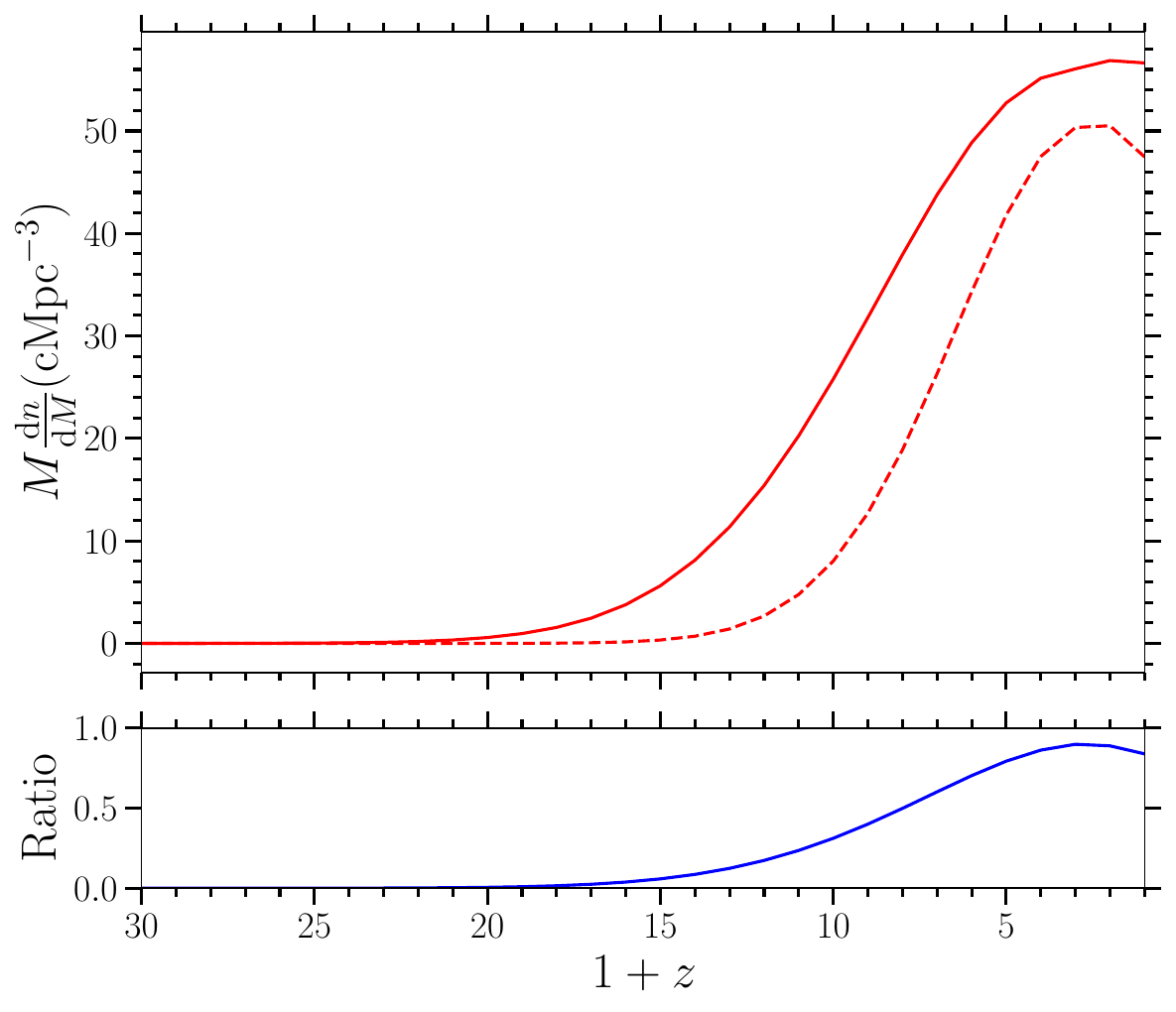}%
\includegraphics[width=0.5\textwidth]{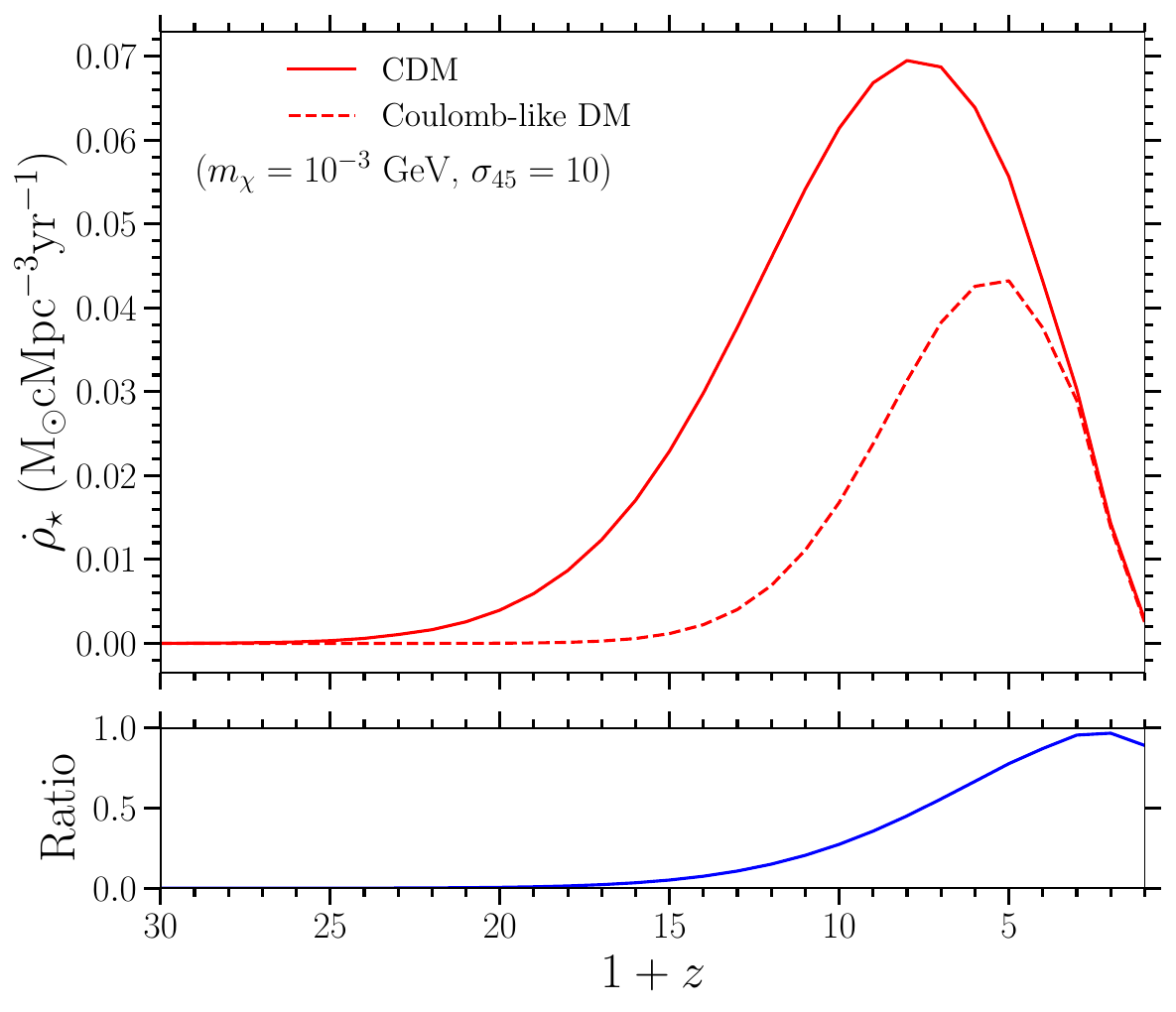}
\caption{\textbf{Left:} the halo mass function (HMF), evaluated at the minimum halo mass for star formation $M_{\mathrm{min}}$ (corresponding to $T^{\mathrm{min}}_\mathrm{vir} = \SI{500}{\kelvin}$), as a function of redshift for standard CDM and IDM. We use the Tinker et al.\ fitting form. \textbf{Right:} the corresponding star formation rate density (SFRD) for the two scenarios. The IDM model is parametrised by a DM particle mass $m_\chi = \SI{1}{\mega\electronvolt}$ and cross-section parameter $\sigma_{45} = 10$ (used in both panels). Both the HMF and the SFRD are suppressed in IDM relative to CDM, reflecting the delayed halo formation that follows from the momentum exchange between DM and baryons.}\label{fig:hmf_sfrd}
\end{figure}
In the case of IDM model, we generate the HMF using a modified version of \verb|CLASS|\footnote{\url{https://github.com/kboddy/class_public/tree/dmeff}} \cite{class,modified_class1,modified_class2,modified_class3} and \verb|Galacticus|\footnote{\url{https://github.com/galacticusorg/galacticus}} \cite{galacticus}, following the procedure described in \cite{Driskell}. We first compute the modified linear matter power spectrum for IDM using the modified \verb|CLASS| code. We then provide this modified power spectrum as an input to \verb|Galacticus| (which is a toolkit for modelling the galaxy formation semi-analytically) which computes $\sigma$ and in turn generates the modified HMF.

The left panel of figure~\ref{fig:hmf_sfrd} compares the Tinker et al.~HMF for standard CDM and IDM (for a representative case of $m_\chi = \SI{0.001}{\giga\electronvolt}$, $\sigma_{45} = 10$ and a halo mass corresponding to minimum virial temperature of $T^{\mathrm{min}}_\mathrm{vir} = \SI{500}{\kelvin}$). As evident from the figure, the HMF suffers a suppression -- due to the momentum transfer between DM and baryons. This suppression in HMF ultimately leads to a suppression in SFRD at all redshifts as shown in the right panel of figure~\ref{fig:hmf_sfrd}.  

In figure~\ref{fig:cooling} we noted how heat exchange leads to excess IGM cooling. However, IDM also affects the star formation history. The primary effect of momentum transfer between DM and baryons is to delay star formation (reflected in the evolution of $\dot{\rho}_{\star}$ with redshift). Since both, Ly$\upalpha$ and X-ray photons, track the star formation, a delayed star formation also delays the Ly$\upalpha$ coupling, Ly$\upalpha$ heating, and X-ray heating (see also discussion in section~\ref{sec:gs}).

\begin{figure}
\centering
\includegraphics[width=0.6\textwidth]{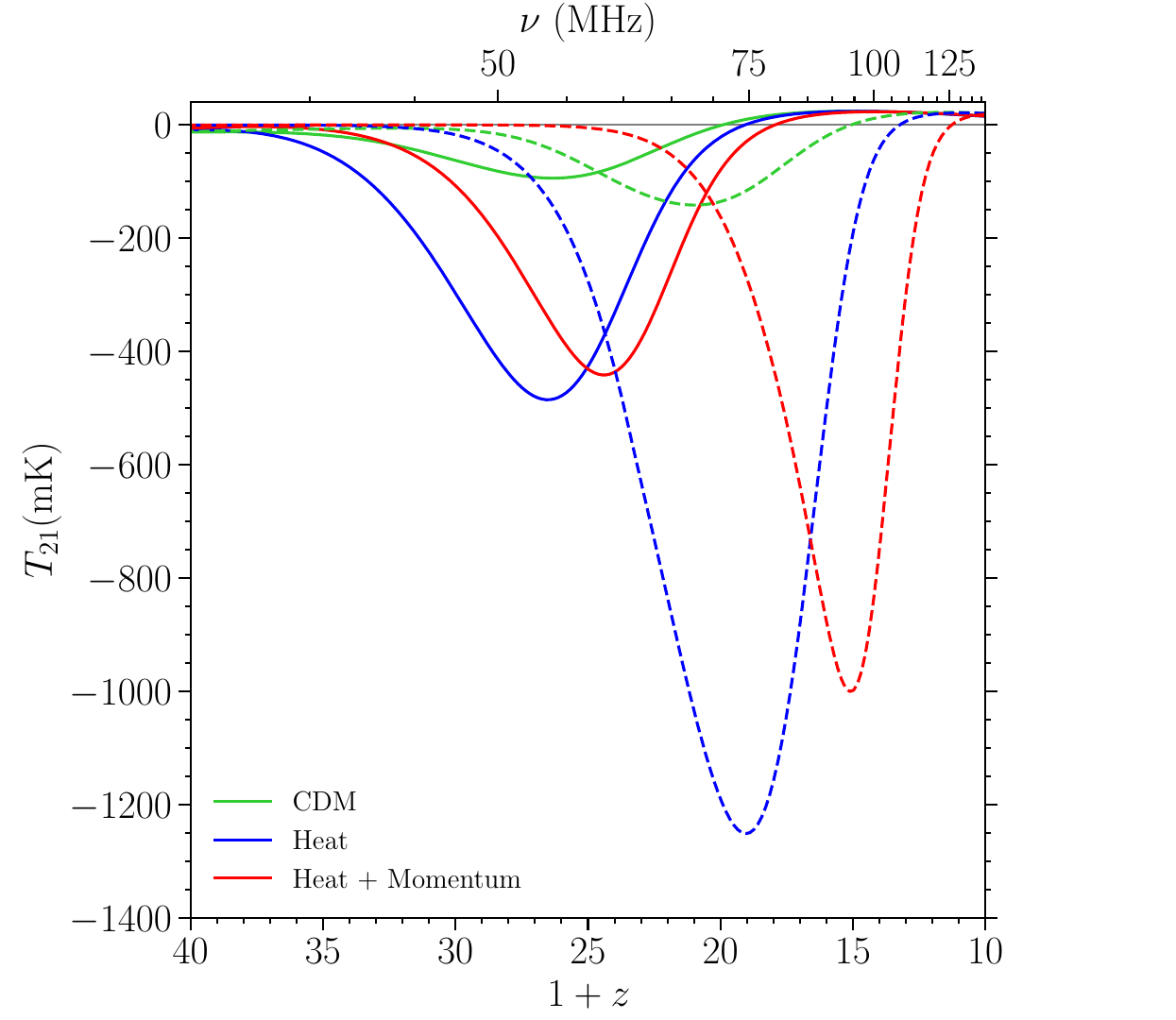}
\caption{Comparison of the global 21-cm signal between baryon-DM heat transfer (blue) vs baryon-DM heat and momentum transfer (red) for two representative cases differing in Ly$\upalpha$ emission strength, star formation threshold, DM mass, and baryon-DM interaction cross-section. For the solid curves the parameter values are $f_{\mathrm{Ly}}=0.1$, $T^{\mathrm{min}}_{\text{vir}} = \SI{500}{\kelvin}$, $m_\chi = \SI{0.32}{\giga\electronvolt} $ and $\sigma_{45} = 1$ while for the dashed curves the parameter values are $f_{\mathrm{Ly}}=1$, $T^{\mathrm{min}}_{\text{vir}} = \SI{e4}{\kelvin}$, $m_\chi = \SI{0.1}{\giga\electronvolt} $ and $\sigma_{45} = 10$. Other astrophysical parameters that are common to both set of curves are $f_{\mathrm{X}} =1, w=1.5$, and $f_{\mathrm{esc}}=0.01$.}\label{fig:astrophysical}
\end{figure}

In figure~\ref{fig:astrophysical}, we demonstrate the impact of this delayed star formation on the global signal for two representative cases that are shown by the set of solid and dashed curves. When only heat exchange is accounted for we typically get a stronger signal (blue curves) compared to the standard CDM model (green curves). When momentum exchange is also included -- for self-consistent model of the IGM -- there is an additional shift along the time as well as temperature axis (red curves). Due to momentum exchange the star formation is suppressed which delays the activation of Ly$\upalpha$ coupling and X-ray heating and correspondingly weakens the 21-cm signal.

\subsection{Charting the parameter space for IDM model}
\begin{figure}
    \centering
    \includegraphics[width=0.5\textwidth]{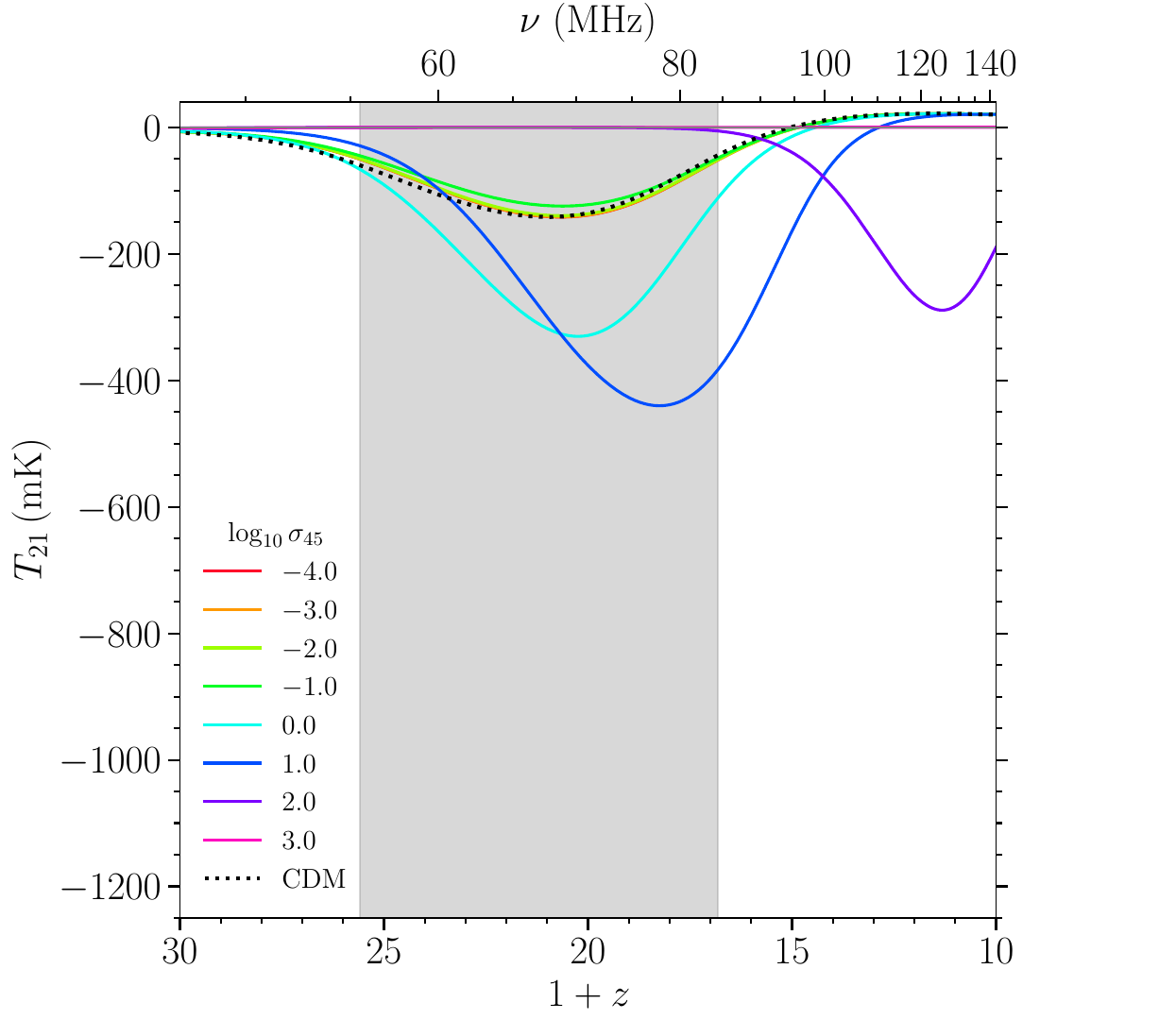}%
    \includegraphics[width=0.5\textwidth]{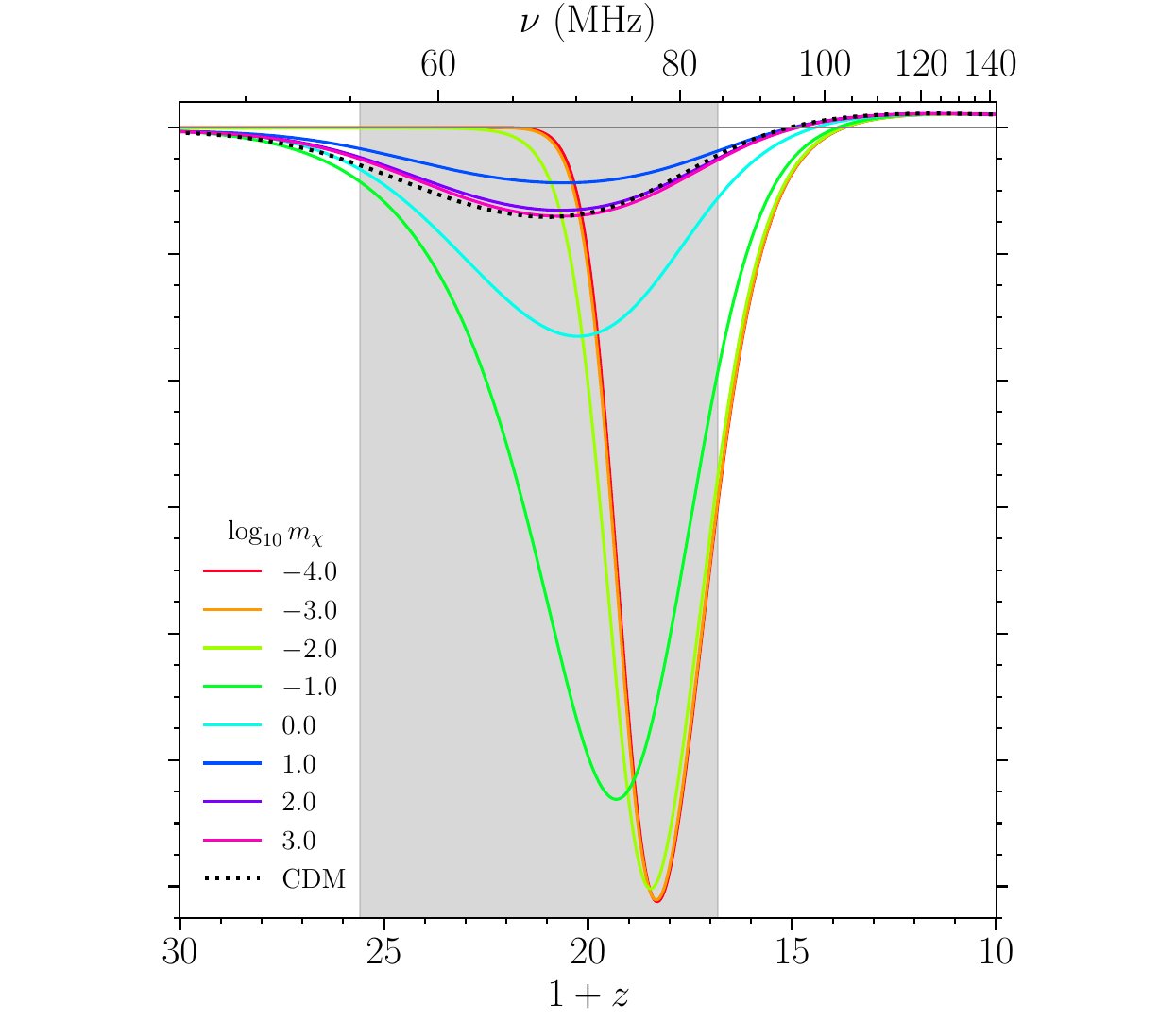}
    \caption{Variation of the global 21-cm signal with the b$\chi$ interaction cross-section (left) and IDM particle mass (right). For both the panels $f_{\mathrm{Ly}}=1, f_{\mathrm{X}} =1, w=1.5, f_{\mathrm{esc}}=0.01$, and $T^{\mathrm{min}}_{\text{vir}} = \SI{e4}{\kelvin}$. For the left panel $m_\chi = \SI{1}{\giga\electronvolt}$ and for the right panel $\sigma_{45} = 1$, where $\sigma_{45}=\sigma_0/(\SI{e-45}{\metre^2})$. The grey shaded region shows the redshift range corresponding to the \textit{SARAS3} data. The black-dotted curve shows the signal for a CDM model with the same astrophysical parameters.}
    \label{fig:combined}
\end{figure}

In figure~\ref{fig:combined}, we demonstrate how the predicted theoretical signal varies with cross-section (left-panel) and with the mass of IDM particle (right-panel) for a representative set of astrophysical parameters. For both panels we fix the astrophysical parameters at $f_{\mathrm{Ly}}=1, f_{\mathrm{X}} =1, w=1.5, f_{\mathrm{esc}} = 0.01$ and $T_{\text{vir}} = \SI{e4}{\kelvin}$. As $\sigma_{45}$ increases from $10^{-1}$ to around 10, heat and momentum exchange increase. As mentioned previously, heat exchange produces cooling, leading to a stronger absorption dip, whereas the effect of momentum exchange is to delay and suppress the SFR which shifts the signal to lower redshift. This is evident from the curves which show a stronger absorption dip as we go from cross-sections $\sigma_{45}$ of $10^{-1}$ to $10$. We note that for very low $\sigma_{45}$ the behaviour is non-monotonic. This behaviour arises because, for the chosen astrophysical parameters, $\sigma_{45}\sim0.1$ corresponds to a regime where drag heating (the second term in eq.~\ref{eq:Qk}) dominates over the cooling term (the first term in eq.~\ref{eq:Qk}) at $z \gtrsim15$. For smaller cross-sections, both terms are suppressed, reducing the overall impact of IDM interactions and causing the signal to approach the CDM prediction (black dotted curve). For larger cross-sections, the cooling term increasingly dominates over drag heating during cosmic dawn and later epochs. Additionally, as we increase $\sigma_{45}$ beyond a value of 10, the SFR is strongly suppressed. This suppression leads to a corresponding reduction in the Ly$\upalpha$ coupling, which weakens the coupling between the spin temperature and the gas temperature. As a result, the depth of the absorption feature decreases, even though the signal continues to shift towards lower redshift values.

The right panel of figure~\ref{fig:combined} shows the variation of the signal with the DM mass $m_\chi$ for a fixed cross-section ($\sigma_{45} = 1$) and other parameters as in the case of left panel. When the DM mass is significantly larger than the characteristic baryonic mass scale, $m_\chi\gg\SI{1}{\giga\electronvolt}$, the drag heating (second term in eq.~\ref{eq:Qk}) dominates over cooling (first term in eq.~\ref{eq:Qk}). This results in a gas temperature that is higher than in the CDM case and in turn weaker signals (blue, purple, and magenta curves). However, for DM mass $m_\chi\ll\SI{10}{\giga\electronvolt}$, the cooling term dominates leading to colder IGM and consequently a stronger 21-cm absorption feature relative to CDM.


An important feature to note from the above figures is how the IDM signals compare with the CDM prediction for identical astrophysical parameters. In the left panel, where $\sigma_{45}$ is varied, we observe that as $\sigma_{45}$ decreases, the signals converge toward the CDM curve (black dotted), as expected since the IDM model reduces to CDM in the limit $\sigma_{45} \to 0$. In the right panel, where $m_\chi$ is varied, the IDM model can produce absorption dips either deeper or shallower than in CDM. The latter occurs when $\SI{1}{\giga\electronvolt} < m_\chi < \SI{100}{\giga\electronvolt}$. In this regime, the first term in eq.~\eqref{eq:Qk} becomes negligible while the second term remains finite, leading to drag heating: baryons are heated rather than cooled, resulting in a reduced absorption dip relative to CDM. On increasing the DM mass beyond $\SI{100}{\giga\electronvolt}$, the interactions become totally inefficient (even the second term drops to zero) and there is effectively no heat exchange between baryons and DM. Thus, for such high masses we see that the absorption dip is very similar to the CDM case. 

Note that the signal response to cross-section or DM mass we show in figure~\ref{fig:combined} is not the general case. For some space of astrophysical parameters it is possible to get a monotonic behaviour \cite{Driskell}. Driskell et al.~\cite{Driskell}, and as have we, show the signals for a representative set of parameters. Our Bayesian analysis pipeline explores a wide range of parameter space and hence consistently accounts for different scenarios.

The effect of astrophysical parameters on the global 21-cm signal is summarised below.
\begin{itemize}
\item $f_{\mathrm{Ly}}$: controls the Ly$\upalpha$ background. Ly$\upalpha$ photons influence the 21-cm signal through two primary roles -- heating and coupling \citep{Mittal_lya}. A higher value of $f_{\mathrm{Ly}}$ indicates a stronger background, resulting in increased heating and enhanced coupling (see the discussion preceding eq.~\ref{eq:falpha_to_heating}). While heating tends to diminish the absorption feature, stronger coupling leads to a more pronounced absorption feature.

\item $f_\mathrm{X}$: controls the X-ray background. As $f_\mathrm{X}$ increases, the X-ray background becomes more intense, leading to a greater X-ray heating. This subsequently raises $T_{\uk}$ and, consequently, $T_{\us}$, resulting in a reduced absorption dip in $T_{21}$.

\item $w$: controls the SED of X-ray emission. For fixed $L_{\mathrm{X}}/\mathrm{SFR}$, decreasing $w$ increases the effective X-ray emission in the energy range $0.2$ to $\SI{30}{\kilo\electronvolt}$ and hence, reduces the signal strength.

\item $f_\mathrm{esc}$: controls the progress of reionization. In general, $f_\mathrm{esc}$ has a small impact on the absorption dip of the signal but a dramatic impact on the emission signal. A low value of $f_\mathrm{esc}$ means low ionizing emissivity which prolongs the reionization epoch. As a result neutral hydrogen population continues to exist for a longer period of time which leads to a late disappearance of the signal. Conversely, a high value of $f_\mathrm{esc}$ leads to a small emission signal and early disappearance of the signal.

\item $T^{\mathrm{min}}_\mathrm{vir}$: controls the threshold for star formation. The impact of minimum virial temperature is much more complex than $f_{\mathrm{Ly}}$ or $f_\mathrm{X}$. However, as shown by Mittal et al.~\cite{Mittal_pbh} increasing $T^{\mathrm{min}}_\mathrm{vir}$ delays the star formation and causes the signal to shift to lower redshifts with a negligible impact on the signal shape.
\end{itemize}

Thus, in total we have seven parameters that describe our global 21-cm signal for IDM model. Table~\ref{tab:parameters} lists these parameters with their brief description.

\begin{table}
\centering
\caption{Free parameters involved in our model of the global 21-cm signal. First column gives the parameter, second column gives the description, third column gives the choice of prior ranges we use for Bayesian inference, and the last column gives the scale of the parameter. For an explanation of priors see text. Note that we do not consider the star formation efficiency ($f_\star$) to be a free parameter and set its value to 0.1 throughout the inference process. Throughout this work we adopt only uniform (i.e. `uninformative' priors).}\label{tab:parameters}
\vspace{2ex}
\begin{tabular}{|llll|}
\hline
Parameter & Description & Prior Range & Scale \\ \hline
\rule{0pt}{2.5ex}$m_\chi$ & Mass of the & & \\
& DM particle (in GeV) & $\left[10^{-4},10^3\right]$ & Log \\
$\sigma_{45}$ & b$\chi$ interaction cross-section  & & \\
    & ($\sigma_0=\sigma_{45}\cdot\SI{e-45}{\metre^2}$) & $\left[10^{-4},10^3\right]$ & Log  \\
$f_\mathrm{Ly}$ & Controls the strength & & \\
    & of Ly$\upalpha$ background & $\left[10^{-2},10^2\right]$ & Log\\
$f_\mathrm{X}$ & Controls the strength & & \\
    & of X-ray background & $\left[10^{-2},10^2\right]$ & Log\\
$w$ & Spectral index of X-ray SED & $[0,3]$ & Linear\\
$f_{\mathrm{esc}}$ & Escape fraction of & & \\
    & ionizing photons & $\left[10^{-3},1\right]$ & Log\\
$T^{\mathrm{min}}_\mathrm{vir}$ & Minimum virial temperature  & & \\
    & of the DM haloes & $\left[10^2,10^6\right]$ & Log \\
\hline
\end{tabular}
\end{table}

\section{Dataset and Inference Procedure}\label{sec:infer}
The \textit{SARAS3} dataset that we use is described by Singh et al.~\cite{SARAS_Detection}. To obtain the data, the team observed the radio intensity spectrum across a range of frequencies between $\nu_{\mathrm{min}}=55.5$ and $\nu_{\mathrm{max}}=\SI{84.4}{\mega\hertz}$ (corresponding to $24.5 \gtrsim z \gtrsim 15.8$). From this, they obtained the residual spectrum by subtracting their best-fitting sky model from the total radio spectrum observed. They modelled the sky as a sixth-order log-log polynomial capturing the galactic and extragalactic foregrounds, ionospheric distortions, systematic errors in calibration, water thermal emission, and antenna efficiency. We will work with the calibrated antenna temperature from the \textit{SARAS} instrument which we will refer to as the \textit{SARAS3} data.

To model the foregrounds-and-systematics (for brevity we just write foregrounds), we use the sixth-order (7 coefficients) log-log polynomial \cite{SARAS_Detection, Bevins_2022}\footnote{Following the analysis of Singh et al.~\cite{SARAS_Detection}, we repeated our joint Bayesian fitting using log-log polynomial foreground models of order $2^\mathrm{nd}$ through $8^{\mathrm{th}}$, each combined with an IDM global-signal component. Consistent with their findings, we find that the $6^{\mathrm{th}}$-order log-log polynomial yields the highest Bayesian evidence and therefore we adopt it as our fiducial foreground model.}
\begin{equation}
T_{\mathrm{fg}}=10^{{\sum_{i=0}^{6}}a_ix^i}\,,\label{t_fs}
\end{equation}
where $x$ is function of frequency normalized to be in range $-1$ to 1
\begin{equation}
x=2\frac{\log_{10}(\nu/\nu_{\mathrm{min}})}{\log_{10}(\nu_{\mathrm{max}}/\nu_{\mathrm{min}})}-1\,.
\end{equation}
Thus, we have a 14-dimensional antenna temperature model, $T_{\mathrm{ant}}=T_{\mathrm{fg}}+T_{21}$, where, as described in the previous section, $T_{21}$ is the 7-dimensional global 21-cm signal for an IDM model, simulated with \texttt{ECHO21}.

Having discussed the data and our sky model, we move to the description of the inference procedure. We use a Bayesian inference procedure to constrain the astrophysical and DM parameters. The procedure, based on Bayes' theorem, allows us to fit a model $\mathcal{M}$ parametrized by $\theta$ to a given dataset $\mathcal{D}$ by updating our prior knowledge of the parameters. Thus, the posterior distribution of parameters can be obtained as
\begin{equation}
\mathrm{P}(\theta|\mathcal{D},\mathcal{M})=\frac{\mathrm{P}(\mathcal{D}|\theta,\mathcal{M})\mathrm{P}(\theta|\mathcal{M})}{\mathrm{P}(\mathcal{D}|\mathcal{M})}\,,
\end{equation}
where $\mathrm{P}(\mathcal{D}|\theta,\mathcal{M})$ is the likelihood, $\mathrm{P}(\theta|\mathcal{M})$ is the prior and $\mathcal{Z}=\mathrm{P}(\mathcal{D}|\mathcal{M})$ is the Bayesian evidence. We use the Python package \texttt{Polychord} \cite{handley1, handley2}, that implements a nested sampling algorithm \cite{Skilling}, for parameter and evidence estimation. We use the default configuration of \texttt{Polychord}; the number of live points is set to be 25 times the model dimensionality, i.e., $25\times14=350$. To post process our posterior chains from \texttt{Polychord}, we use \texttt{anesthetic} \cite{anesthetic} for creating corner plots and \texttt{fgivenx} \cite{fgivenx} for creating functional posterior distribution.

Bayesian evidence can be used to quantify the preference (model quality) amongst competing models. Additional Bayesian statistics such as Kullback-Leibler (KL) divergence \cite{kl} and Gaussian model dimensionality (GMD) are also useful in assessing the constraining power of the data. The KL divergence ($\mathcal{D_\text{KL}}$) can be used to measure the information gain from prior to posterior and thus it provides a natural metric for how strongly the data constrain the model parameters. In particular, $\mathcal{D_\text{KL}}$ can be interpreted as the logarithm of the effective compression of the prior volume induced by the likelihood, with larger values corresponding to posteriors that occupy a smaller fraction of the prior volume. On the other hand, the GMD ($d_\mathrm{G}$), which is the posterior variance of the log likelihood, quantifies the effective number of dimensions being constrained by the data \cite{Handley_2019}.

To determine the constraining power of the data on the astrophysics of the first galaxies and the properties of IDM, $\theta_{21} = (m_\chi, \sigma_{45}, \ldots, T^{\mathrm{min}}_{\mathrm{vir}})$, we need to marginalize over the foreground parameters $\theta_{\mathrm{fg}}=\{a_i\}$ before calculating the KL divergence and GMD. The foreground parameters do not carry information about the astrophysical processes at high redshift and are `nuisance' parameters for us. Computing statistics for selected parameters involves a high-dimensional integral of the posterior distribution over the nuisance parameters. We use the Real NVP implementation from the publicly-available Python package \texttt{margarine} to learn from the posterior samples the density $\mathrm{P}(\theta_{21}|\theta_{\mathrm{fg}}, \mathcal{M}, \mathcal{D})$ and evaluate the marginal KL divergence and GMD \cite{margarine, Bevins_2023}. Real-valued non-volume preserving (real NVP) transformation is a type of normalizing flow method for density estimation and is appropriate for structured high-dimensional distributions such as ours \cite{Laurent_2017}. For the architecture of our neural network, we choose 6 affine coupling transformations, 16 hidden layers in each coupling layer, and 64 neurons for each hidden layer. 

Due to the uncertainty in the astrophysical processes at high redshifts, we work with uninformative, or equivalently uniform, prior distributions (either in log space or linear space) to perform our Bayesian analysis. In order to consider a wide range of possible signal models, we choose broad priors on all parameters around the conventionally accepted values in literature that are either motivated by observations or state-of-the-art simulations (see ref.~\cite{Mittal_echo}, and references therein).  Following the above discussion, we have chosen our priors on the free parameters that are summarised in table~\ref{tab:parameters}. For the foregrounds (eq.~\ref{t_fs}) we adopt a range $-10$ to $10$ in linear scale as the priors for all $a_i$'s \cite{SARAS_Detection, Bevins_2022}.

We define our likelihood function as follows. Let $T_{\mathrm{ant}}^{\mathcal{D}}=T_{\mathrm{ant}}^{\mathcal{D}}(z)$ and $T_{\mathrm{ant}}^{\mathcal{M}}=T_{\mathrm{ant}}^{\mathcal{M}}(\theta,z)$ represent the data and the model-predicted values, respectively of the antenna temperature at redshift $z$. We define a Gaussian likelihood as
\begin{equation}
\mathcal{L}=\mathrm{P}\left(T_{\mathrm{ant}}^{\mathcal{D}}|\theta\right) = \prod_{i=1}^{470} \frac{1}{\sqrt{2\pi\varepsilon_i^2}} \exp\left[ -\frac{\left(T_{\mathrm{ant}}^{\mathcal{D}}-T_{\mathrm{ant}}^{\mathcal{M}}\right)_i^2}{2 \varepsilon_i^2}\right]\\,
\end{equation}
where $i$ runs over the 470 frequency bins (equivalently redshift bins) in the \textit{SARAS3} data. We adopt a diagonal covariance matrix with frequency-dependent variance: $\varepsilon_i$ is the per-bin measurement noise reported by Singh et al.\ (figure 1(c) of \cite{SARAS_Detection}), which we treat as the dominant source of uncertainty in the data.

Bayesian inference over our parameter space requires a large number of likelihood evaluations, each of which calls the global 21-cm signal model. Computing the signal on the fly for every likelihood call is prohibitively expensive, so we instead pre-compute the signal on a regular grid of parameters (see table~\ref{tab:parameters}) and obtain the signal for any desired parameter set by linear interpolation. To validate this, we held out a subset of grid points and, for each, computed the RMS of the difference of interpolated and true signal across a wide redshift band. The distribution of per-point RMS errors has a mean of $\SI{2.5}{\milli\kelvin}$ and a 95th percentile of $\SI{8.4}{\milli\kelvin}$, which is well below the \textit{SARAS3} noise level ($\SI{213}{\milli\kelvin}$).

As a validation of our inference pipeline, we performed signal-recovery tests on mock antenna-temperature data injected with a known IDM signal at two noise levels: $\SI{25}{\milli\kelvin}$ and $\SI{250}{\milli\kelvin}$ (the latter comparable to \textit{SARAS3}). The injected signal is recovered cleanly at low noise but not at high noise. The pipeline therefore behaves as expected, and the noise level of \textit{SARAS3} is the limiting factor for parameter inference. Details are given in appendix~\ref{sec:pipetest}.

\section{Results and Discussion}\label{sec:rnd}
We now present the constraints on our astrophysical and DM model parameters jointly fit with a sixth-order log-log polynomial against the \textit{SARAS3} data following the inference scheme described in the previous section.

\begin{figure}
\centering
\includegraphics[width=1\textwidth]{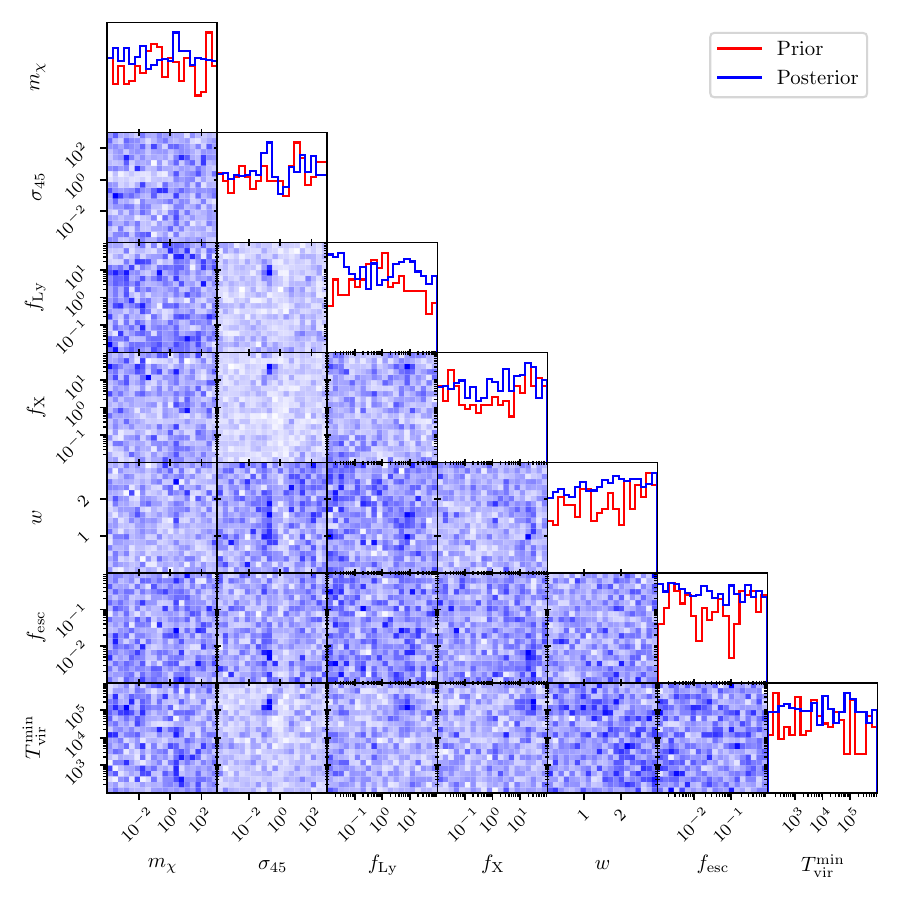}
\caption{One- and two-dimensional marginalized posterior distributions for the signal parameters in the IDM scenario, obtained by fitting a sixth-order log-log polynomial foreground model together with the global 21-cm signal to the \textit{SARAS3} data; for the foreground parameter posteriors, see figure~\ref{fig:all_posterior_idm}. Neither individual parameters nor pairs of parameters are constrained by the data, although the functional posterior does place constraints on signal amplitude (see figure~\ref{fig:signal_poste_idm}). See table~\ref{tab:parameters} for parameter definitions and prior ranges.}\label{fig:mcmc_results}
\end{figure}
Figure~\ref{fig:mcmc_results} shows our marginalized two-dimensional and one-dimensional posterior distributions for the signal parameters only. Figure~\ref{fig:all_posterior_idm} in appendix~\ref{sec:joint} shows the posterior for the foreground parameters and the 2D posteriors for foreground and signal parameters. Posteriors of our signal parameters cover the entire range of priors, indicating that the \textit{SARAS3} data is not good enough to constrain the values of the individual parameters. We can quantify this with the marginal KL divergence, which for the 7D space of signal parameters comes out to be $\mathcal{D}_{\mathrm{KL}}\approx0.007$. A small KL divergence implies that prior and posterior volumes are essentially the same, implying no information gain. Similarly, the marginal GMD is $d_{\mathrm{G}}\approx0.01$, which again implies that effectively no signal parameters are being constrained. 

\begin{figure}
\centering
\includegraphics[width=0.8\textwidth]{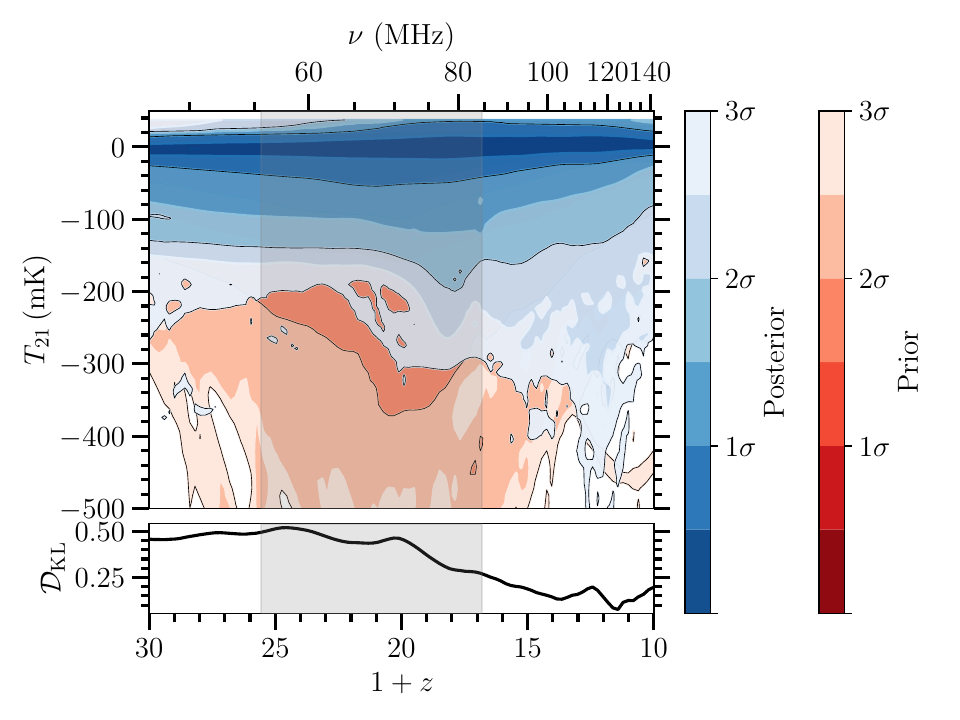}
\caption{Functional prior (red) and posterior (blue) samples of the global 21-cm signal as a function of redshift, for the IDM scenario. The grey shaded region indicates the redshift band covered by \textit{SARAS3}. \textbf{Bottom:} KL divergence between the functional prior and posterior at each redshift. The signal is most strongly constrained at $z = 23.6$, where $\mathcal{D}_{\mathrm{KL}}$ peaks; at this redshift the data place a $3\sigma$ lower bound of $T_{21}(z=23.6) \gtrsim \SI{-277.6}{\milli\kelvin}$ on the signal amplitude.}\label{fig:signal_poste_idm}
\end{figure}
While we do not obtain strong constraints on individual parameter (or even pair of parameters), it is possible to rule out strong signals. In figure~\ref{fig:signal_poste_idm} we show our functional distribution of posterior and prior.\footnote{Note that because of the non-linear dependence of the signal on parameters, a uniform sampling in parameter space does not translate to a uniform sampling in signal space.}\,The red regions show the prior samples that cover signals as deep as $\sim\SI{-2000}{\milli\kelvin}$ within the \textit{SARAS3} band (for visual clarity we clip the $y$ axis at $-\SI{500}{\milli\kelvin}$). The blue regions show the posterior samples, which are contracted to a somewhat smaller space by the data. The bottom panel shows the KL divergence in the functional space at each redshift bin. Within the \textit{SARAS3} band, the signal is most strongly constrained at $z=23.6$ ($\nu\approx\SI{57.7}{\mega\hertz}$) with a KL divergence of $0.52$. The 21-cm signal stronger than $\SI{-277.6}{\milli\kelvin}$ at this redshift is ruled at $3\sigma$ level. Stated differently, \textit{SARAS}3 data puts the following constraint: 
\begin{equation*}
T_{21}(z=23.6)\gtrsim \SI{-277.6}{\milli\kelvin}\,,
\end{equation*}
at $3\sigma$ level on models of the 21-cm signal for IDM. This is expected given the RMS value of our residuals.

\begin{figure}
\centering
\includegraphics[width=1\linewidth]{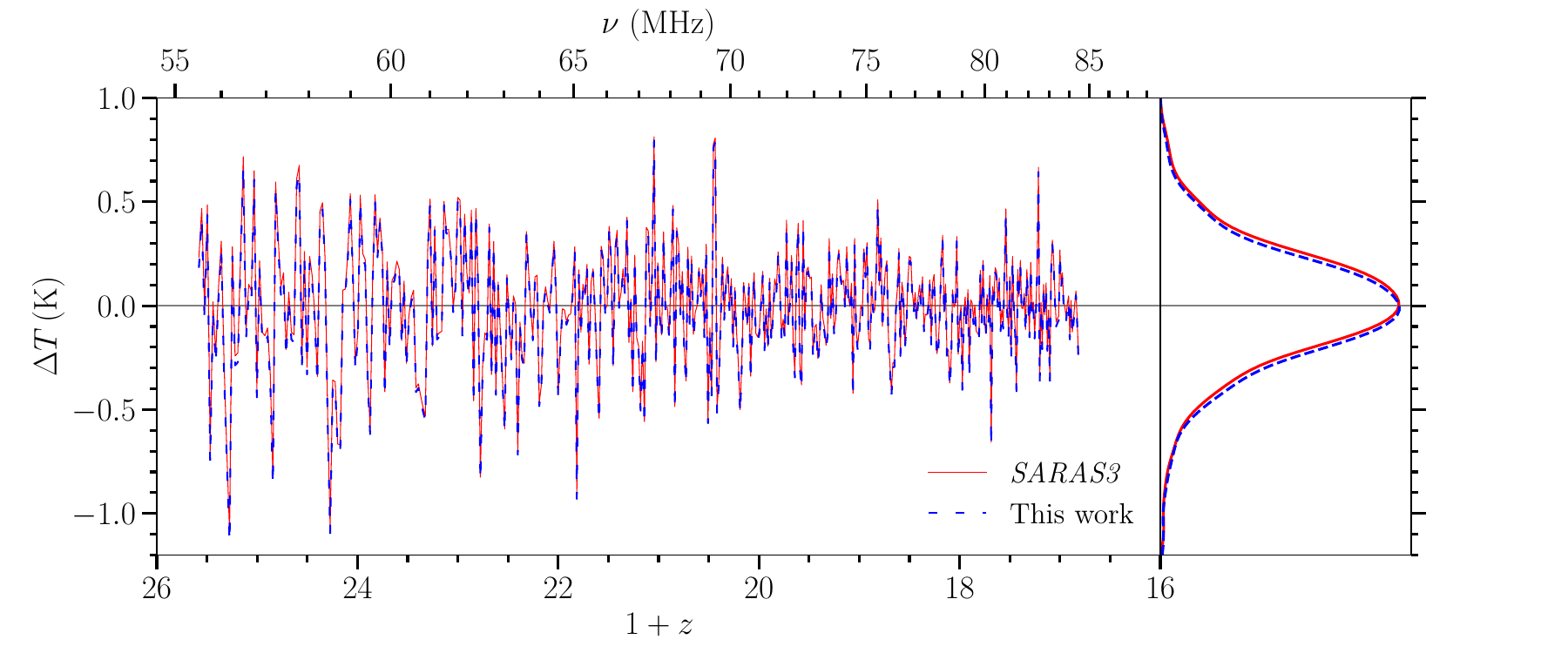}
\caption{Residuals from the original \textit{SARAS3} analysis (red, solid) and from this work (blue, dashed). The latter are obtained by subtracting our best-fit foreground, the weighted median of the posterior of the $a_i$ coefficients, from the antenna data. The RMS values are 213 and $\SI{215}{\milli\kelvin}$ respectively, consistent with the published \textit{SARAS3} result. \textbf{Right:} distributions of the two residual sets.}\label{fig:residual}
\end{figure}
Unlike the signal parameters, foreground parameters are much more strongly constrained. Accordingly, we can subtract our best-fitting foregrounds model (corresponding to weighted-median of the posterior samples) from the \textit{SARAS3} antenna data to obtain the residuals. Accordingly, the weighted RMS (weighted by $\varepsilon_i^{-2}$) value of this residual is $\SI{215}{\milli\kelvin}$. In figure~\ref{fig:residual}, we compare our residuals with that reported by \textit{SARAS3} (figure~1(b) of \cite{SARAS_Detection}), which they obtained with MCMC analysis. The right-hand side panel shows the distribution of residuals. The overlap seen in the figure and the small difference between the residuals ($\sim\SI{2}{\milli\kelvin}$) indicates that our foreground modelling is consistent with that reported by Singh et al.~\cite{SARAS_Detection} (see also \cite{Bevins_2022}).\footnote{As a consistency check, we have also done fitting without the 21-cm signal, i.e., only foregrounds and find the weighted RMS of the residual to be $\approx213\,$mK, in agreement with \cite{SARAS_Detection}.}

Our results are in qualitative agreement with the galaxy-dominated radio models results of Bevins et al.~\cite{Bevins_2022}. In their analysis, radio-background models give rise to a strong global 21-cm signal, while in our case IDM models play an analogous role. In both scenarios, strong 21-cm signals in the \textit{SARSA3} observation band are in tension with the data. Consequently, our posteriors weakly disfavour high $f_{\mathrm{Ly}}$, low $f_{\mathrm{X}}$, and low $T^{\mathrm{min}}_{\mathrm{vir}}$ (though only marginally so, given that the marginalized posteriors remain close to the prior), consistent with the conclusions of Bevins et al., albeit expressed in a different parametrization.

To assess if there is a preference of IDM model over the CDM model by \textit{SARAS3}, we compute the Bayesian evidence for the CDM scenario and compare it with that of IDM scenario. In appendix~\ref{app:cdm_results} we present the Bayesian analysis results for a pure CDM model. The log Bayesian evidences for IDM and CDM are $-3342.40$ and $-3342.93$, respectively. Consequently, the log Bayes factor is
\begin{equation*}
\ln B = \ln \mathcal{Z}_{\mathrm{IDM}} - \ln \mathcal{Z}_{\mathrm{CDM}} = 0.53\,,
\end{equation*}
so that $B\approx1.7$, corresponding to betting odds in favour of the IDM model of 1.7:1. According to the Jeffreys scale, this corresponds to an inconclusive evidence, indicating no strong preference of one DM model over the other.\\

Our wide astrophysical priors implied by the high uncertainty in the astrophysics of high-redshift Universe, the nature of data and the flexibility of foreground model prevents us from making any strong claims on astrophysical or cosmological parameters. Stronger bounds on these parameters should be obtainable from wider band data of global 21-cm experiments such as the \textit{REACH} and by narrowing down the priors on astrophysical parameters with other observational datasets. Datasets that could be utilized to reduce the high-redshift astrophysical uncertainties include cosmic X-ray background \cite{Harrison_2016}, 21-cm power spectrum \cite{Abdurashidova_2023}, UV luminosity functions \cite{Chakraborty_2026, Souradeep_2025}. Some recent works have also explored the possibility of combining datasets to constrain the high-redshift astrophysics \cite{Pochinda_2024, Omer_2025, Dhandha_2025, Sims_2025}.

\section{Conclusions}\label{sec:conc}
In this work, we addressed three connected questions. First, whether a Coulomb-like interacting dark matter (IDM) model is consistent with the \textit{SARAS3} data; second, what constraints the data place on the IDM parameters; and third, whether the data show any preference for IDM over cold dark matter (CDM). The IDM model considered here has two key effects on the 21-cm signal. The first is non-adiabatic cooling of baryons through heat transfer to the DM. The second is a delay in cosmic star formation due to suppression of halo formation. This delay shifts the absorption trough to lower redshifts and reduces its depth, since Ly$\upalpha$ coupling, Ly$\upalpha$ heating, and X-ray heating are all postponed accordingly.

For inference, we simultaneously fit a global 21-cm signal with a foregrounds model against the \textit{SARAS3} antenna data. We use the publicly-available Python code \texttt{ECHO21} for simulating the global signal, self-consistently accounting for both the cooling as well as the star-formation effects on the intergalactic medium. Our model for the 21-cm signal is described by seven parameters, namely DM particle mass $m_\chi$, the interaction cross-section parameter $\sigma_0$, scale factors for the Ly$\upalpha$ intensity $f_{\mathrm{Ly}}$, and the X-ray intensity $f_{\mathrm{X}}$, power-law index of X-ray emissivity $w$, escape fraction of ionizing photons $f_{\mathrm{esc}}$ and the minimum virial temperature of star-forming dark matter haloes $T^{\mathrm{min}}_{\mathrm{vir}}$. For foregrounds, we adopted a sixth-order log-log polynomial following the previous works. We did a nested-sampling-enabled Bayesian analysis for the inference.

The main conclusions of this work are as follows:
\begin{enumerate}
\item Due to the large uncertainty in the data and the degeneracy between the foregrounds and the cosmological signal, the data do not provide meaningful constraints on the signal parameters. This is evident from the information gain and effective dimensionality metrics: the KL divergence is very small (0.007), and the Gaussian model dimensionality is also negligible (0.01). Together, these results indicate that the data add minimal constraining power beyond the prior for the signal parameters.

\item However, the \textit{SARAS3} data are not entirely devoid of information. By translating the parameter posteriors into functional space posteriors of the global 21-cm signal, we find that the strongest constraint from the \textit{SARAS3} data occurs at redshift $z = 23.6$ ($\nu \approx \SI{57.7}{\mega\hertz}$), where the KL divergence reaches $0.52$. At this redshift, the data rule out signals stronger than $\SI{-277.6}{\milli\kelvin}$ at the $3\sigma$ level.

\item To assess if there is any preference of IDM over CDM, we compared the Bayesian evidence for an IDM model with that for the CDM model. We obtained a Bayes factor of $\approx1.7$ for IDM relative to CDM. Based on Jeffrey's scale this is inconclusive evidence, and hence we cannot claim a preference for IDM over CDM.
\end{enumerate}

The 21-cm cosmic dawn absorption signal is subject to astrophysical uncertainties, which make precise predictions of its depth and redshift range uncertain. Further improvements in narrowing the range of uncertainty on the astrophysical priors by combining datasets from complementary probes of cosmic dawn could help partially pin down the expected signal range. Our work highlights the importance of a joint foregrounds and signal analysis and the fact that null-like dataset are not devoid of information about astrophysics and cosmology.

With this work we release our updated Python package \texttt{ECHO21}(\url{https://github.com/shikharmittal04/echo21.git}), with a feature to model the global signal for Coulomb-like interacting dark matter.

\acknowledgments
We thank Andrew Benson, Jiten Dhandha, Mansi Dhuria, Trey Driskell, Dragan Huterer, Nicholas Kern, Girish Kulkarni, and Vikram Rentala for useful discussions. PB acknowledges support from the U.S. Department of Energy under Contract No.~DE-SC0019193. SM is supported by the ERC (UKRI guaranteed) research grant EP/Y02916X/1. HTJB acknowledges support from the Kavli Institute for Cosmology Cambridge and the Kavli Foundation. This work has been performed as part of the DAE-STFC Technology and Skills Programme ‘Building Indo-UK collaborations towards the Square Kilometre Array’. This work used the DiRAC Data Intensive service (CSD3) at the University of Cambridge, managed by the University of Cambridge University Information Services on behalf of the STFC DiRAC HPC Facility (\href{https://dirac.ac.uk/}{www.dirac.ac.uk}). The DiRAC component of CSD3 at Cambridge was funded by BEIS, UKRI and STFC capital funding and STFC operations grants. DiRAC is part of the UKRI Digital Research Infrastructure.

\bibliographystyle{JHEP}
\bibliography{biblio}

\appendix

\section{Test of pipeline with mock data}\label{sec:pipetest}
In the main text we worked with \textit{SARAS3} data and performed inference for an interacting dark matter model. In this appendix we test our inference pipeline on mock data. Assuming a \textit{SARAS}-like, minimally chromatic antenna (so chromatic distortions to the foreground are neglected), we construct mock antenna data comprising a foreground, a 21-cm signal, and a noise component. For the foreground we adopt the simple empirical form
\begin{equation}
T_{\mathrm{fg}} = T_0\left(\frac{\nu}{\nu_0}\right)^{\alpha}\,,
\end{equation}
where $T_0=\SI{335.4}{\kelvin}, \nu_0=\SI{150}{\mega\hertz}$, and $\alpha=-2.8$. We add an IDM 21-cm signal generated with parameters $m_\chi=\SI{1}{\giga\electronvolt}, \sigma_{45}=1, f_{\mathrm{Ly}}=1, f_{\mathrm{X}}=1, w=1.5, f_{\mathrm{esc}}=0.01$, and $T^{\mathrm{min}}_{\mathrm{vir}}=\SI{e4}{\kelvin}$. This signal reaches an absorption minimum of $\SI{-340}{\milli\kelvin}$ at $\nu=\SI{70}{\mega\hertz}$ ($z\approx19$). Finally, we add white Gaussian noise; we consider two cases with standard deviation: $25$ and $\SI{250}{\milli\kelvin}$.

\begin{figure}
\centering
\includegraphics[width=1\linewidth]{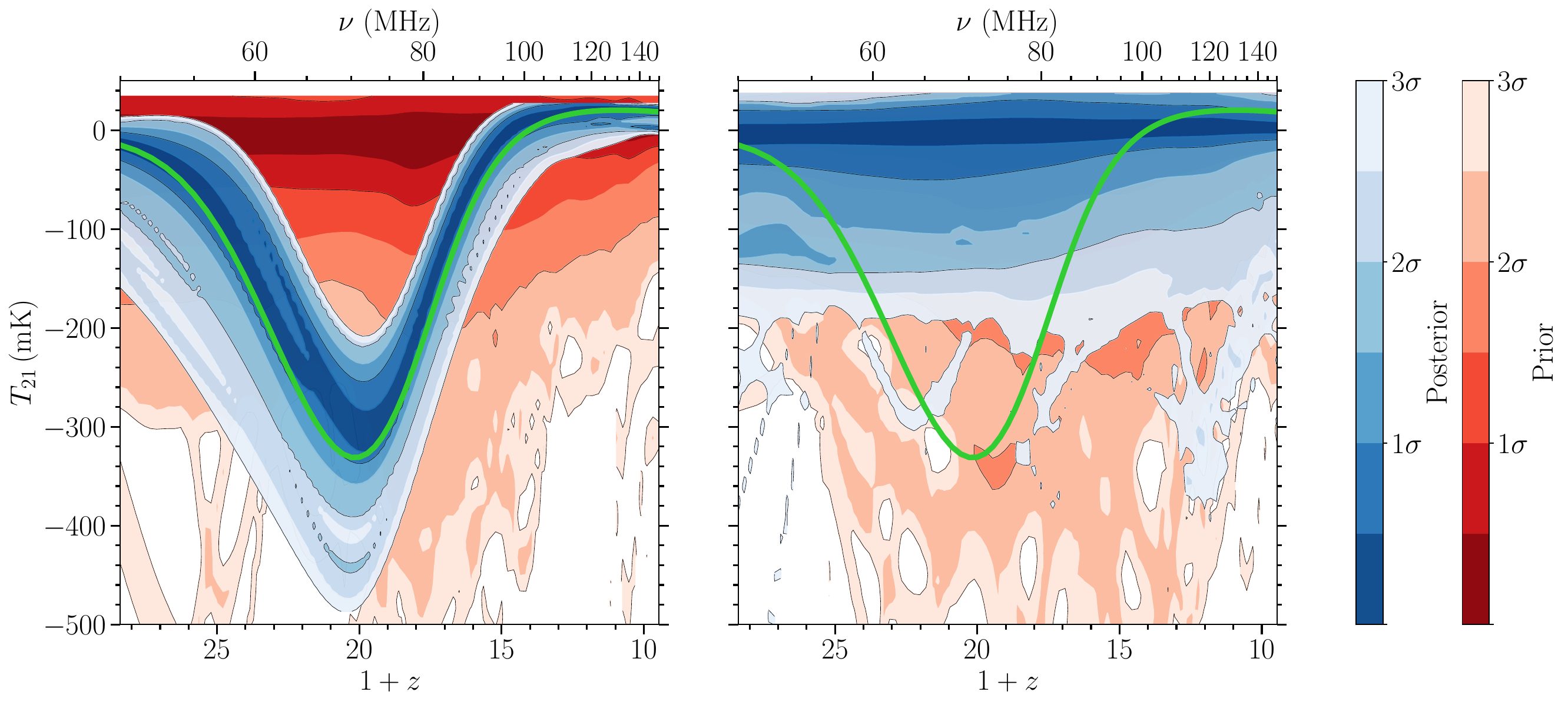}
\caption{Functional prior (red) and posterior (blue) samples of the global 21-cm signal as a function of redshift, for the mock antenna temperature data injected with an IDM 21-cm signal (green curve). \textbf{Left:} mock data with noise level of $\SI{25}{\milli\kelvin}$. \textbf{Right:} mock data with noise level of $\SI{250}{\milli\kelvin}$ (comparable to \textit{SARAS3}). The injected signal is well recovered at low noise (left) but not for \textit{SARAS3}-level noise (right).}\label{fig:mock}
\end{figure}

Following the inference procedure developed in section~\ref{sec:infer}, we perform a joint fit of the foreground and signal to the mock data. Figure~\ref{fig:mock} shows the functional priors and posteriors, with the true injected signal overlaid as a green curve. As expected, the signal recovered from the low-noise mock data (25\,mK, left) is in good agreement with the injected signal, whereas the high-noise case ($\SI{250}{\milli\kelvin}$, comparable to \textit{SARAS3}; right) shows no such recovery.

\section{Joint posterior distribution for IDM model}\label{sec:joint}
\begin{figure}
    \centering
    \includegraphics[width=0.5\linewidth]{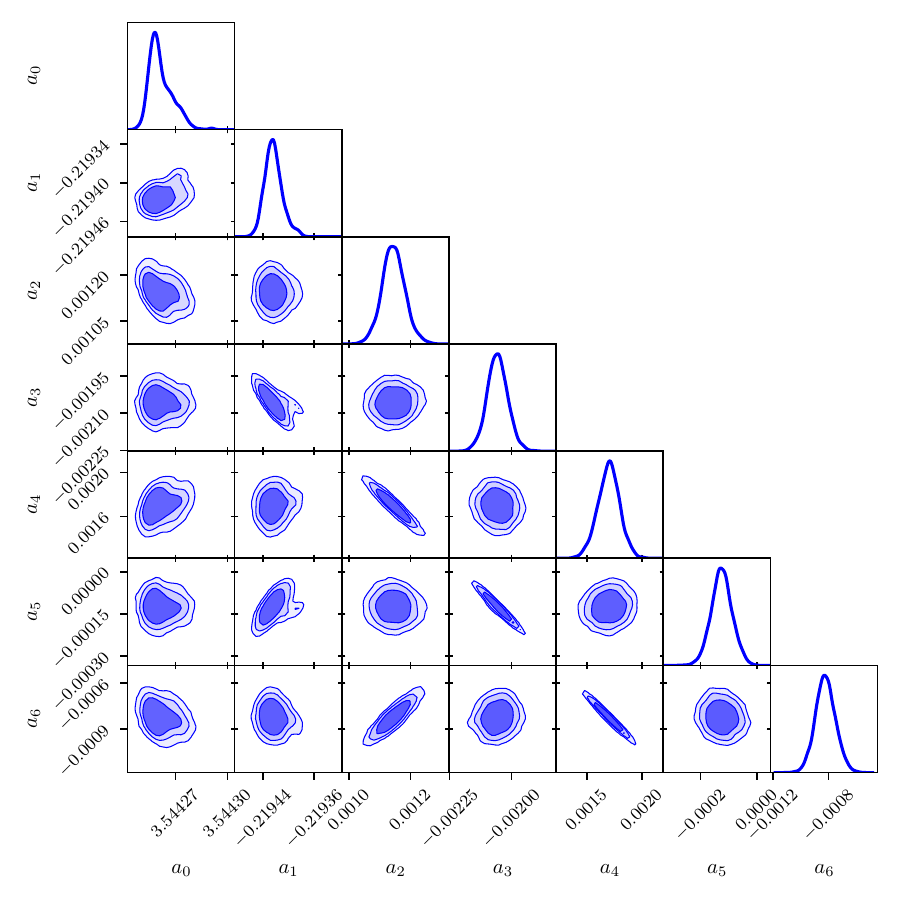}%
    \includegraphics[width=0.5\linewidth]{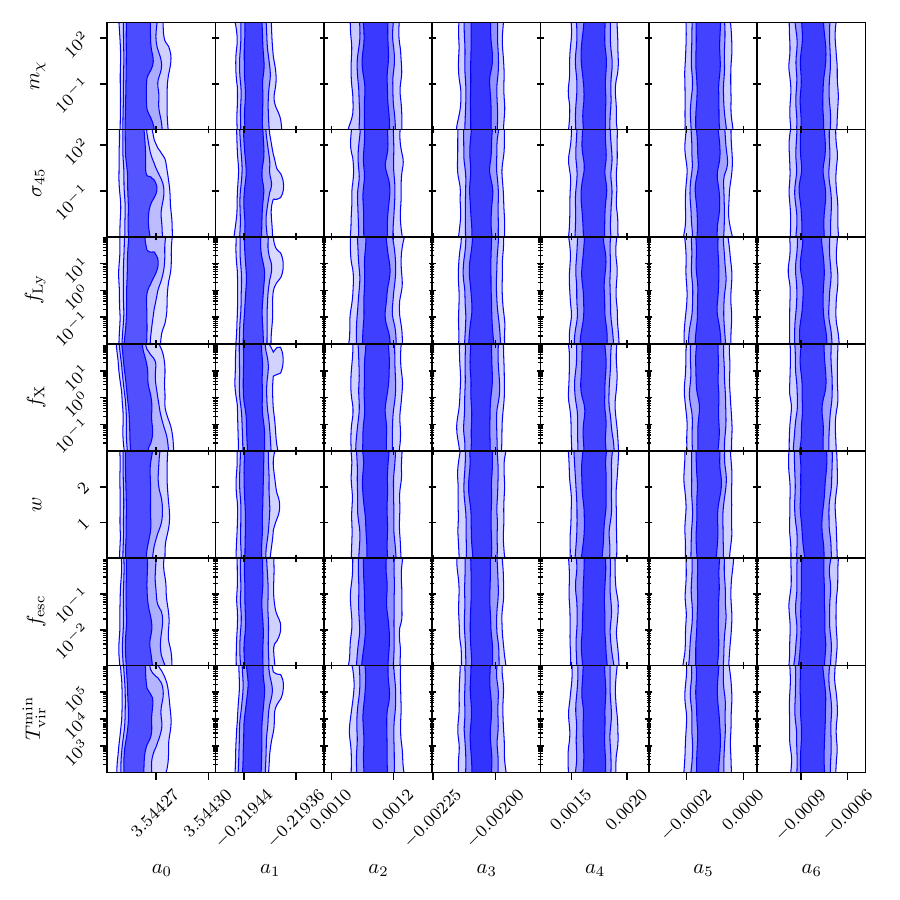}
    \caption{Marginalized posteriors from the IDM fit to the \textit{SARAS3} data. \textbf{Left:} 1D and 2D posteriors for the foreground polynomial coefficients $a_i$ (sixth-order log-log polynomial). \textbf{Right:} joint posterior of the signal and foreground parameters, showing both the tight constraints on the foreground coefficients and the broad, near-prior distributions of the signal parameters. See table~\ref{tab:parameters} for the signal parameter definitions and prior ranges.}\label{fig:all_posterior_idm}
\end{figure}

Figure~\ref{fig:all_posterior_idm} shows the posteriors from our IDM fit to the \textit{SARAS3} data: foreground parameters alone (left) and the joint posterior of foreground and signal parameters (right). As detailed in section~\ref{sec:infer}, we adopt a sixth-order log-log polynomial for the foregrounds. The weighted-median values of the polynomial coefficients are:
\begin{align*}
a_0 &= +3.54\\
a_1 &=-0.22\\
a_2 &= +1.14\times10^{-3}\\
a_3 &= \num{-2.10e-3}\\
a_4 &= +\num{1.70e-3}\\
a_5 &= \num{-1.25e-4}\\
a_6 &= \num{-8.24e-4}
\end{align*}
The uncertainty on each $a_i$ is $\mathcal{O}(10^{-5})$ or smaller, so we do not list individual error bars. The posteriors for the polynomial coefficients are tightly constrained, whereas the signal-parameter posteriors occupy essentially their full prior range, consistent with the data constraining the smooth foreground but not the signal.

\section{Implications of \textit{SARAS3} data for cold dark matter model}\label{app:cdm_results}
In this appendix, we present the constraints on astrophysical parameters for the standard cold dark matter (CDM) scenario from \textit{SARAS3} data. The global 21-cm signal in CDM is governed by Ly$\upalpha$ coupling and heating, X-ray heating, and reionization. The corresponding emissions, i.e., Ly$\upalpha$, X-ray, and ionizing photons, all track the instantaneous star formation rate density. Accordingly, the model is parametrized by $f_\mathrm{Ly},\, f_\mathrm{X},\, w,\, f_{\mathrm{esc}}$, and $T^{\mathrm{min}}_\mathrm{vir}$ \cite{Mittal_echo}. For our Bayesian analysis, we use the same priors listed in table~\ref{tab:parameters}, with \( m_\chi \) and \( \sigma_{45} \) excluded since the dark matter is non-interacting in this case.

\begin{figure}
\centering
\includegraphics[width=1\textwidth]{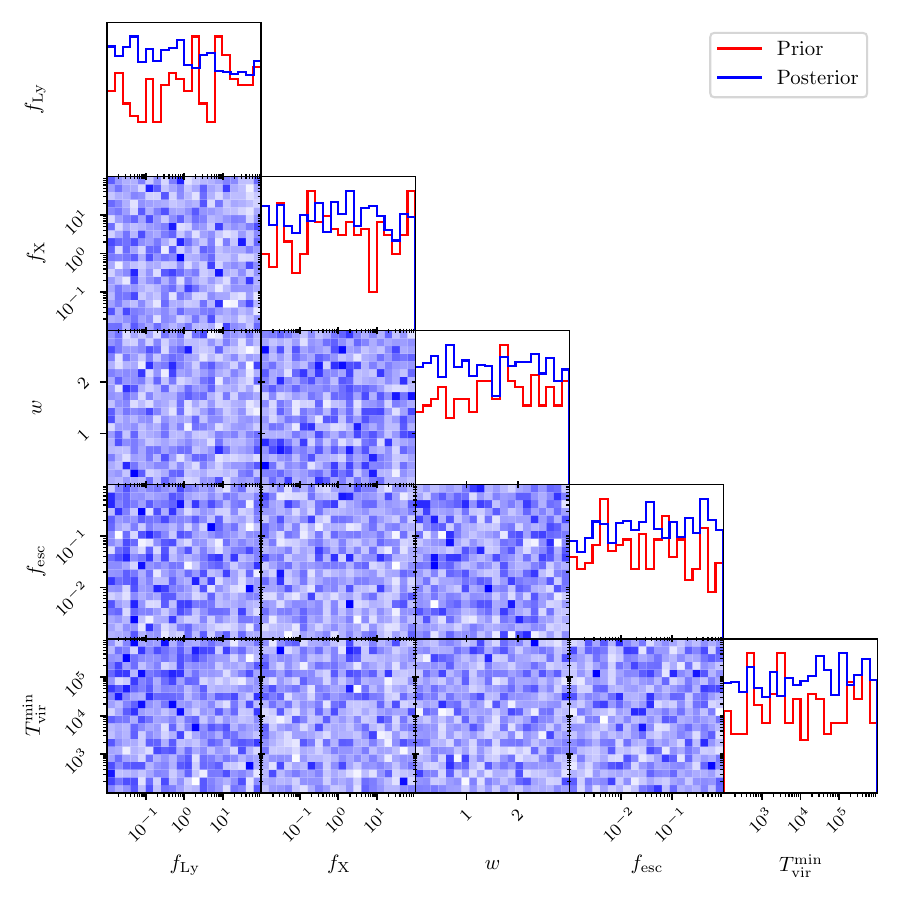}
\caption{One- and two-dimensional marginalized posterior distributions for the signal parameters in the CDM scenario, obtained by fitting a sixth-order log-log polynomial foreground model together with the global 21-cm signal to the \textit{SARAS3} data. The 1D posteriors are flat across their priors, indicating that the data do not constrain the signal parameters in this case (marginal $\mathcal{D}_{\mathrm{KL}}\approx 3\times10^{-3}$ and marginal $d_{\mathrm{G}} \approx 5\times10^{-3}$). See table~\ref{tab:parameters} for parameter definitions and prior ranges.}\label{fig:mcmc_results_cdm}
\end{figure}

In figure~\ref{fig:mcmc_results_cdm} we show the marginalized two-dimensional and one-dimensional posterior distributions. As in the case of IDM, the 1D posteriors for the signal parameters are flat across the prior range, indicating that the data do not constrain these parameters. Quantitatively, the marginal KL divergence is $\mathcal{D}_{\mathrm{KL}}\approx 3\times10^{-3}$ and marginal GMD is $d_{\mathrm{G}} \approx 5\times10^{-3}$, both consistent with negligible contraction of the prior volume.

\begin{figure}
\centering
\includegraphics[width=0.8\textwidth]{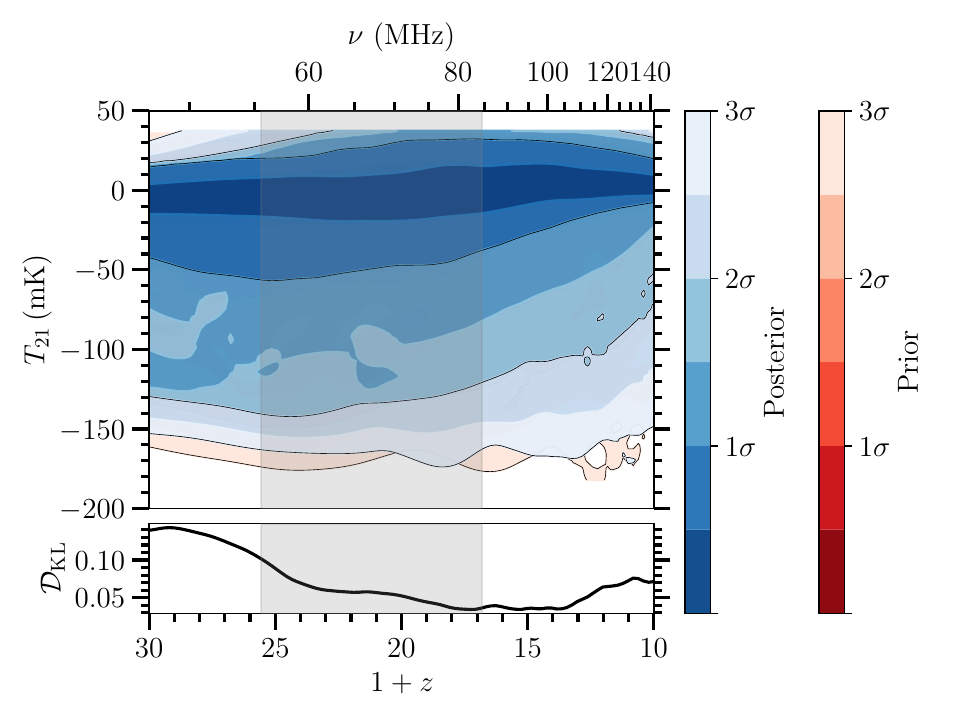}
\caption{Functional prior (red) and posterior (blue) samples of the global 21-cm signal as a function of redshift, for the CDM scenario. The grey shaded region indicates the redshift band covered by \textit{SARAS3}. \textbf{Bottom:} KL divergence between the functional prior and posterior at each redshift. The posterior nearly overlaps the entire prior, reflecting the fact that the data do not constrain the signal in this case (compare with the IDM case in figure~\ref{fig:signal_poste_idm}).}\label{fig:signal_poste_cdm}
\end{figure}
However, unlike the IDM case, \textit{SARAS3} does not place any constraint on the signal magnitude. In the standard CDM scenario, the prior contains signals only as deep as $\sim\SI{-180}{\milli\kelvin}$ -- well below the RMS of the residuals (\SI{213}{\milli\kelvin}) -- whereas the IDM prior extended to depths of $\sim\SI{-2000}{\milli\kelvin}$, which the data could exclude. Accordingly, we expect negligible contraction of the functional-space prior in the CDM case. Figure~\ref{fig:signal_poste_cdm} shows the functional posteriors (blue) overlaid on the functional priors (red); as the figure makes clear, the posterior nearly covers the entire prior space.

\end{document}